\documentclass[11pt]{article}

\usepackage[top = 1in,bottom = 1in, textwidth = 6.5in, textheight=9.1in]{geometry}

\usepackage{amsmath}
\usepackage{graphicx}
\usepackage{latexsym}
\usepackage{setspace}
\usepackage{xcolor}
\usepackage{amssymb}
\usepackage{accents}
\usepackage{tabularx}
\usepackage{fancyhdr}
\usepackage{verbatim}
\usepackage{multirow}
\usepackage{framed}
\usepackage[square,sort,comma,numbers]{natbib}
\usepackage{url}
\usepackage[final]{pdfpages}
\usepackage{float, subfig}
\usepackage{enumitem}
\usepackage{mathtools}
\usepackage{mathrsfs}
\usepackage{amsfonts}
\usepackage{listings}
\usepackage{amsthm}
\usepackage{grffile}
\usepackage{etoolbox}
\usepackage{sidecap}
\usepackage{diagbox}
\usepackage[utf8]{inputenc}
\setboolean{@twoside}{false}
\newcommand{\tcb}{\textcolor{black}}

\def\qed{\hfill{\(\vcenter{\hrule height1pt \hbox{\vrule width1pt height5pt
				\kern5pt \vrule width1pt} \hrule height1pt}\)} \medskip}

\renewcommand{\ss}{\smallskip}
\newcommand{\ms}{\medskip}

\allowdisplaybreaks

\renewcommand{\underline}{\underaccent{\bar}}

\begin{document}
	\baselineskip = 0.25in
	\begin{center}
		\begin{large}
			\begin{bf}
				
				Analyzing Client Behavior in a Syringe Exchange Program \ms
				
			\end{bf}
		\end{large}
		\begin{small}
			\today \ss\\
		\end{small}
		Haoxiang Yang\(^1\), Yue Hu\(^2\), David P.\ Morton\(^1\) \ss\\
		\begin{scriptsize}
			\setstretch{1}
			\(^1\)Department of Industrial Engineering and Management Sciences, Northwestern University, Evanston IL, 60208.\\
			haoxiangyang2019@u.northwestern.edu, david.morton@northwestern.edu\\
			\(^2\)Decision, Risk, and Operations, Columbia Business School, New York NY, 10027.\\
			yhu22@gsb.columbia.edu\\
		\end{scriptsize}
	\end{center}
	
	\begin{abstract}
		Multiple syringe exchange programs serve the Chicago metropolitan area, providing support for drug users to help prevent infectious diseases. Using data from one program over a ten-year period, we study the behavior of its clients, focusing on the temporal process governing their visits to service locations and \tcb{on} their demographics. We construct a phase-type distribution with an affine relationship between model parameters and features of an individual client. The phase-type distribution governs inter-arrival times between reoccurring visits of each client and is informed by characteristics of a client including age, gender, ethnicity, and more. The inter-arrival time model is a sub-model in a simulation that we construct for the larger system\tcb{, which} allows us to provide a personalized prediction regarding the client's time-to-return to a service location so that better intervention decisions can be made \textcolor{black}{with the help of simulation.}\\
		\newline
		\textbf{Keywords:} syringe exchange program, personalized prediction, phase-type distribution, discrete-event simulation
	\end{abstract}
	
	\section{Introduction}\label{sec:ProbDesc}
	Three major agencies provide syringe exchange programs (SEPs) in the Chicago metropolitan area: Community Outreach Intervention Projects (COIP), Chicago Recovery Alliance (CRA), and Test Positive Aware Network (TPAN). Each agency offers equipment and educational services and conducts research on drug users. With the goal of supporting persons who inject drugs (PWIDs) and helping prevent the spread of infectious diseases, they provide services including street outreach, counseling and training for preventing HIV and hepatitis C, case management for persons living with HIV, assistance in entering treatment for substance use, and HIV medical, mental, and pharmacy care. Their locations include storefronts and mobile vans, which may operate according to a flexible schedule. 
	\textcolor{black}{The benefits of an SEP are twofold. First, multiple studies have shown that SEPs are effective in reducing risk behavior such as sharing syringes \cite{bluthenthal2007higher, bluthenthal2000effect, braine2004long, Holtzman09, huo2007needle}, thereby lowering rates of HIV and hepatitis~C transmission. Second, higher utilization of an SEP provides PWIDs with more opportunities to learn about treatment programs, which further reduces drug use~\cite{huo2006Cessation,desimone2005needle}.}
	
	\textcolor{black}{This paper focuses on one of the SEPs in Chicago, and we refer to program participants as~\emph{clients.} Service locations, including storefronts and mobile vans, accept used syringes and, in exchange, provide clients with new syringes along with other devices that help prevent the spread of disease, such as condoms, cookers, purified water, and bleach. On their first visit to a service location, clients are asked to take a voluntary survey involving demographic information and the nature of their drug use (frequency, types of drugs, etc.), which we detail in Section~\ref{sec:Data}. Once a client is established in the system, the SEP keeps a record of frequency of drug use, health condition, and general living condition. During a visit, an SEP employee will have a personal conversation with a client, e.g., about recent life changes, employment status, and family situation. The SEP will further provide the client with information to help with health issues, and seek to introduce the client to drug treatment programs. Evidence has shown that this type of personal interaction can help clients obtain peer support to recover from substance addiction \cite{clarke2016harmreduction,hay2017influence,kidorf2008expanding}.}
	
	\textcolor{black}{Nationwide, about 681,000 Americans aged 12 years or older reported using heroin in 2013 \cite{NSDUH2013}, and the number of reported users grew every year from 2007 to 2013, with new users growing about 70\% from 2002 to 2013. The volume of heroin seized by officials, and the number of heroin overdoses, both grew over the same period in Chicago \cite{NSDUH2013}. The contrast between the significant growth in the use of heroin and slightly lower use of services at the target SEP (see Section~\ref{sec:Data}) motivates our study. In order to promote its services and tailor them to individual clients, the agency needs a better understanding of client behavior in using the SEP. }
	
	\textcolor{black}{The arrival process of clients to SEP sites is key to understanding their use of SEP services, and we have data on arrival times and locations over a ten-year period. Our main focus is to develop a {\em contextual} understanding of the arrival process; i.e., we want to understand inter-arrival times given {\em features} of an individual client obtained, in part, from the voluntary demographic survey. Our data suggest some clients ``establish care" with the SEP, returning consistently, while others use SEP services once and never return. We seek to model and understand the ``life cycle" of an individual client from initiation, reoccurring visits, and termination with the program. Through interviews with SEP staff, we learned that ethnicity, gender, age, geographic location, and drug history affect how a client will use SEP services. For example, African Americans are less active in SEP programs because they tend to prefer not to disclose their drug habits, and because African Americans who use drugs tend to snort rather than inject. A client's interaction with the SEP is affected by life events; e.g., a person who moves farther from a service location may visit less frequently. These are snapshots of a pervasive phenomenon that suggest many factors can influence client behavior when using SEP services.} 
	
	\tcb{To address these issues, we propose a model for how a client engages with the SEP using three integrated sub-models for initiation, reoccurring visits, and termination. The latter two sub-models are fully integrated and involve a phase-type distribution, which may be viewed as a continuous-time Markov chain with hidden states. To build a contextual model, we express the Markov chain's parameters as a function of client features using linear and logistic regression. Analysis with our model can help the SEP understand the importance of different features and hence estimate the distribution governing a specific client's next arrival time. This, in turn, can help SEP staff better allocate limited resources to improve the program's effectiveness. Armed with a probability distribution for the timing of a client's next visit, SEP staff can be alerted when a specific client has not used SEP services for an unusual period of time, which may point to risky drug-use behavior. SEP employees can then contact the client (e.g., via a text message), or dispatch a mobile van to specific locations and message nearby clients. Using simulation, we illustrate the potential value of using our model in this manner.  }
	
	\tcb{Modeling arrival processes plays a key role in many application domains; see~\cite{lakshmi2013application} for a review of relevant literature in healthcare. Homogeneous and nonhomogeneous Poisson processes are widely used to model an arrival process~\cite{aksin2007modern, fomundam2007survey, govil1999queueing}, but do not address our primary goal, i.e., to provide insights regarding a client's return time based on individual predictors associated with that client. In principle, we could partition clients into different categories based on their features and fit such arrival processes based on these categories. However, such an approach scales poorly given the number of features we consider. }
	
	\tcb{As we will discuss, our data on inter-arrival times of clients to SEP sites have heavier tails than that of an exponential distribution, and this is true even when we develop models that condition on sub-populations of the clients. This may arise for two related reasons: (i)~the ``establishing care'' nature of some clients discussed above, and (ii)~clients effectively transitioning between hidden states, which capture active and passive engagement with the SEP. The three sub-model approach that we propose allows us to capture these effects, and its contextual nature captures heterogeneity. We use a negative binomial model for initiation, which is consistent with an over-dispersed mixture of Poisson distributions that captures heterogeneity. We use a two-state continuous-time Markov chain model with unobservable states to capture reoccurring visits and termination. These two integrated sub-models allow us to reasonably represent issues~(i) and~(ii) and their conditional transition- and system-exit probabilities allow us to capture heterogeneity.}
	
	\tcb{Hidden Markov chain models have been used to model scenarios in which an agent transitions between a modest number of states, each associated with certain patterns of behavior.  Paddock et al.~\cite{paddock2012epidemiological} construct a Markov chain model to understand trajectories of a marijuana user with the goal of using simulation to assess alternative treatment and prevention policies. Liu et al.~\cite{liu2015efficient} learn a continuous-time hidden Markov chain model for disease progression in glaucoma and  Alzheimer's disease. Chehrazi et al.~\cite[Appendix~C]{chehrazi2019dynamic} discuss using a two-state Markov chain to model the repayment behavior associated with delinquent credit card accounts, in which the two states represent high and low repayment rates, with the goal of directing credit collection efforts. Hidden Markov chains enable modelers to capture plausible, but unobservable, transitions of an agent. Our approach can further enhance such models by allowing the parameters of the Markov chain to depend on agent-specific features using regression. Given requisite data, the types of models just sketched could benefit from our approach, increasing model fidelity and insights from analysis, by linking agent heterogeneity to the Markov chain model. A technical challenge that we address in this paper---at least for the family of serial Coxian models that we detail---involves parameter estimation when using regression to map the features of an agent to the Markov chain's parameters.}
	
	We first provide, in Section~\ref{sec:Data}, an overview of the demographic and arrival data collected by the SEP between 2005 and 2014. In Section~\ref{sec:Model}, we provide details of the derivation of the three sub-models: initiation, reoccurring visits, and termination, which together capture individual features associated with clients. We present our computational techniques and results in Section~\ref{sec:experiment}, with model validation \tcb{and an example of active intervention that our model can recommend.} We conclude the paper in Section~\ref{sec:Conclusions} by summarizing the model and insights from our analysis.
	
	\section{Description of the Data} \label{sec:Data}
	The data from the SEP consist of results from a survey and records of individual syringe exchange transactions. The transaction data were collected from January 2001 to November 2014 with 139,488 entries. The survey data were collected between July 2005 and November 2014 with 6,843 surveys. Each survey entry corresponds to a unique client. When a new client arrives at a service site, the client is assigned a unique study number (henceforth, client ID) and is asked to complete an enrollment survey. This client ID is then used throughout the client's sojourn in the system. 
	
	We use the data from July 2005 to November 2014 because the surveys are aligned consistently with the transaction records in this period. After removing incomplete and contaminated records, we combine the transaction and survey data to obtain a merged dataset with 63,960 entries, each with 50 data fields, which we detail in the following section.  
	\subsection{Survey Data}
	\tcb{The survey contains 31 questions, which yield 33 predictors,} covering basic demographic information, \tcb{ZIP code of residence,} and the client's drug use habits. These predictors are categorized as follows:
	\begin{itemize}
		\setlength\itemsep{0.1em}
		\item Basic personal information: age, ethnicity, gender;
		\item Length of time \textcolor{black}{using/injecting} drugs;
		\item Frequency of \textcolor{black}{using/injecting} drugs \textcolor{black}{in the past 30 days;}
		\item Frequency of reusing \textcolor{black}{one's own} syringes \textcolor{black}{in the past 30 days;} 
		\item Frequency of sharing syringes \textcolor{black}{with others in the past 30 days;} 
		\item Involvement in group injection \textcolor{black}{in the past 30 days;} 
		\item Sources of syringes;
		\item Types of drugs used;
		\item Drug treatment program participation \textcolor{black}{in the past six months;} 
		\item \textcolor{black}{Reasons and frequency of being in the area with an SEP location in the past 30 days.} 
	\end{itemize}
	Among 5,903 clients in our merged dataset, 4,101 are male, 1,800 are female, and two are transgender. The demographics of these clients are summarized in Tables~\ref{table:ethnicity} and~\ref{table:age}.\\
	\begin{table}[H]
		\centering
		\begin{tabular}{| l | r | r |}
			\hline
			Ethnicity & Number of clients & Percentage (\%) \\ \hline
			White & 3,045 & 51.58 \\ \hline
			African American & 1,384 & 23.45 \\ \hline
			Puerto Rican & 873 & 14.78 \\ \hline
			Mexican & 364 & 6.17 \\ \hline
			Other Latino & 72 & 1.22 \\ \hline
			Other & 165 & 2.80 \\ \hline
			Total & 5,903 & 100 \\ \hline
		\end{tabular}
		\caption{Ethnicity of clients}
		\label{table:ethnicity}
	\end{table}
	\begin{table}[H]
		\centering
		\begin{tabular}{| l | r | r |}
			\hline
			Age (years) & Number of clients & Percentage (\%) \\ \hline
			\(<\)15 & 1 & 0.02 \\ \hline
			\(16 - 30\) & 2,602 & 44.08 \\ \hline
			\(31 - 45\) & 2,108 & 35.71 \\ \hline
			\(46 - 60\) & 1,111 & 18.82 \\ \hline
			\(>60\) & 81 & 1.37 \\ \hline
			Total & 5,903 & 100 \\ \hline				
		\end{tabular}
		\caption{Age of clients}
		\label{table:age}
	\end{table}
	
	The average age of first drug use among all clients is 23.5 years. About 67.9\% of clients inject drugs daily, and 81.7\% injected drugs more than 20 out of 30 days before taking the survey. For all reasonable responses to the survey question regarding injection frequency (defined as at most 10 injections per day), the average number of injections per day is 2.77. About 82.6\% of clients reported that they did not use someone else's syringe \textcolor{black}{in the past 30 days}, and 77.9\% reported that no one used their syringes \textcolor{black}{over the same time period}. Among 986 clients who injected with syringes others had used \textcolor{black}{in the past 30 days}, 558 shared with their spouse, 53 shared with a family member, 350 shared with a friend, 24 shared with an acquaintance, and 11 shared with a stranger. About 72.1\% of the clients did not share cookers, cotton or water during injection, and 8.6\% of the clients stated that they injected drugs in a shooting gallery within the 30 days before they began using the syringe exchange service. About 69.2\% of the clients were in the neighborhood of the SEP location where they took the survey more than 20 days in a typical 30-day month, 25.1\% of whom lived in the neighborhood with an SEP location, 54.5\% of whom had come mainly to buy drugs, 2.0\% of whom had come mainly to exchange syringes, and 4.6\% of whom had come to visit friends. \textcolor{black}{The voluntary nature of the survey could lead to a non-response bias in the results; see, for example, the illustrated cases and discussion in \cite{locker1981bias, Cheung2017phsurvey}. Because every SEP transaction is labeled with a client ID, we can compute the fraction of clients for whom we have a completed survey, which is 98.6\%. Discussions with SEP staff confirmed that although the survey is voluntary, almost every new client takes the survey upon the recommendation of SEP staff, and so the overall effect of non-response bias is reasonably small in our study.}
	
	\subsection{Transaction Data}\label{subsec:transaction}
	The transaction data include details of syringe exchanges that occurred in multiple SEP storefronts and on mobile van routes. During each transaction, SEP employees record the client ID, number of syringes exchanged, number of other preventive devices \tcb{distributed}, size of the group coming with the client, and type(s) of health education material given to the client. 
	In between July 2005 and November 2014, the SEP distributed 3,647,384 syringes, 160,895 male and female condoms, and 63,667 sets of educational material.
	The mean size of the group coming with the client during a single transaction \textcolor{black}{was} 1.89 with a standard deviation of 2.37. The mean number of syringes exchanged in one transaction \textcolor{black}{was} 57.03 with a standard deviation of 110.88. 
	\tcb{Our discussions with the SEP suggested that there are ample staff to process clients' visits, and that the SEP always had enough syringes and rarely ran out of other drug-injection equipment. Thus we see the level of censoring in the demand data as minimal.}
	
	\begin{figure}
		\centering
		\includegraphics[width=\textwidth*9/10]{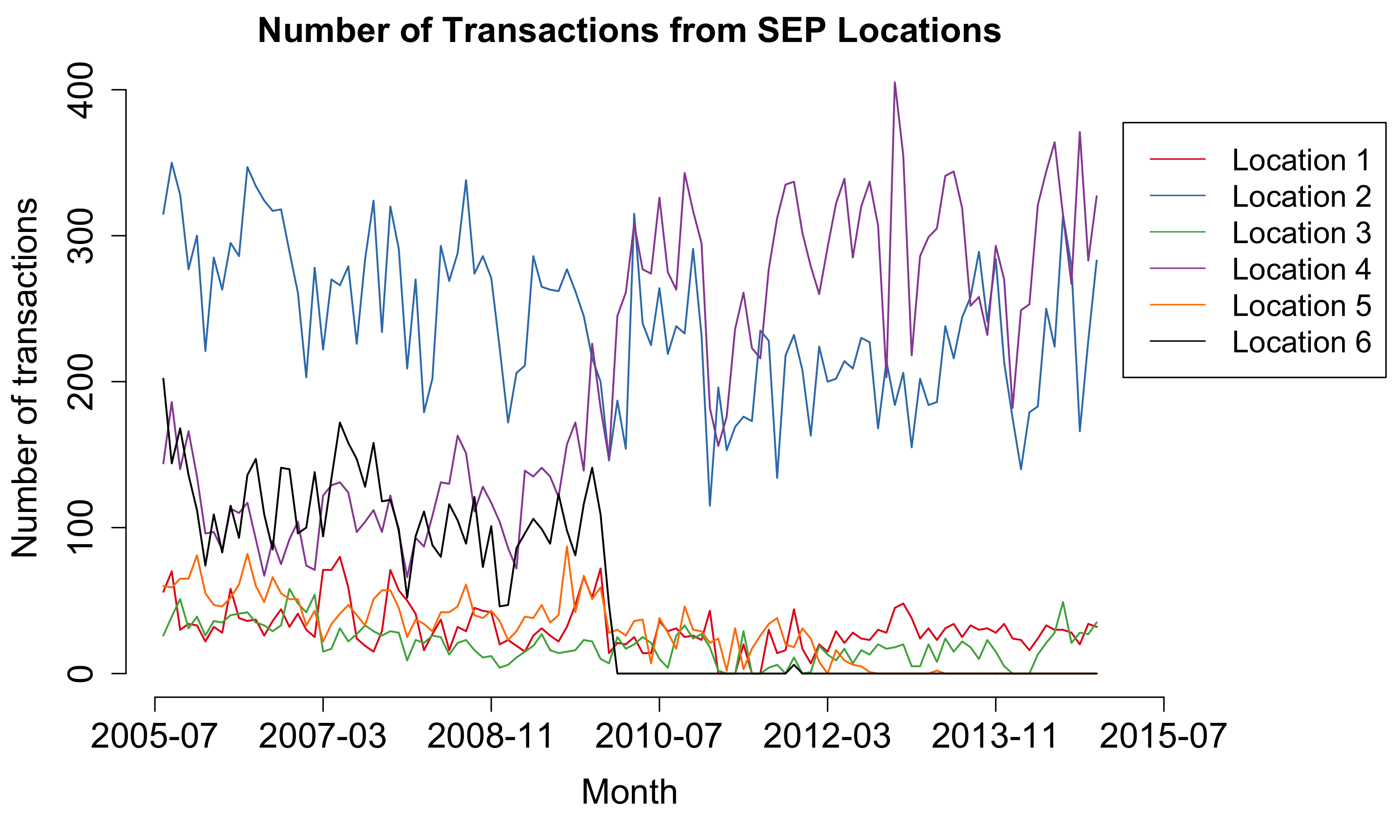}
		\caption{The time series of transactions from \tcb{six} SEP service locations}
		\label{fig:TimeLoc}
	\end{figure}
	
\begin{figure}
		\centering
		\includegraphics[width=\textwidth*9/10]{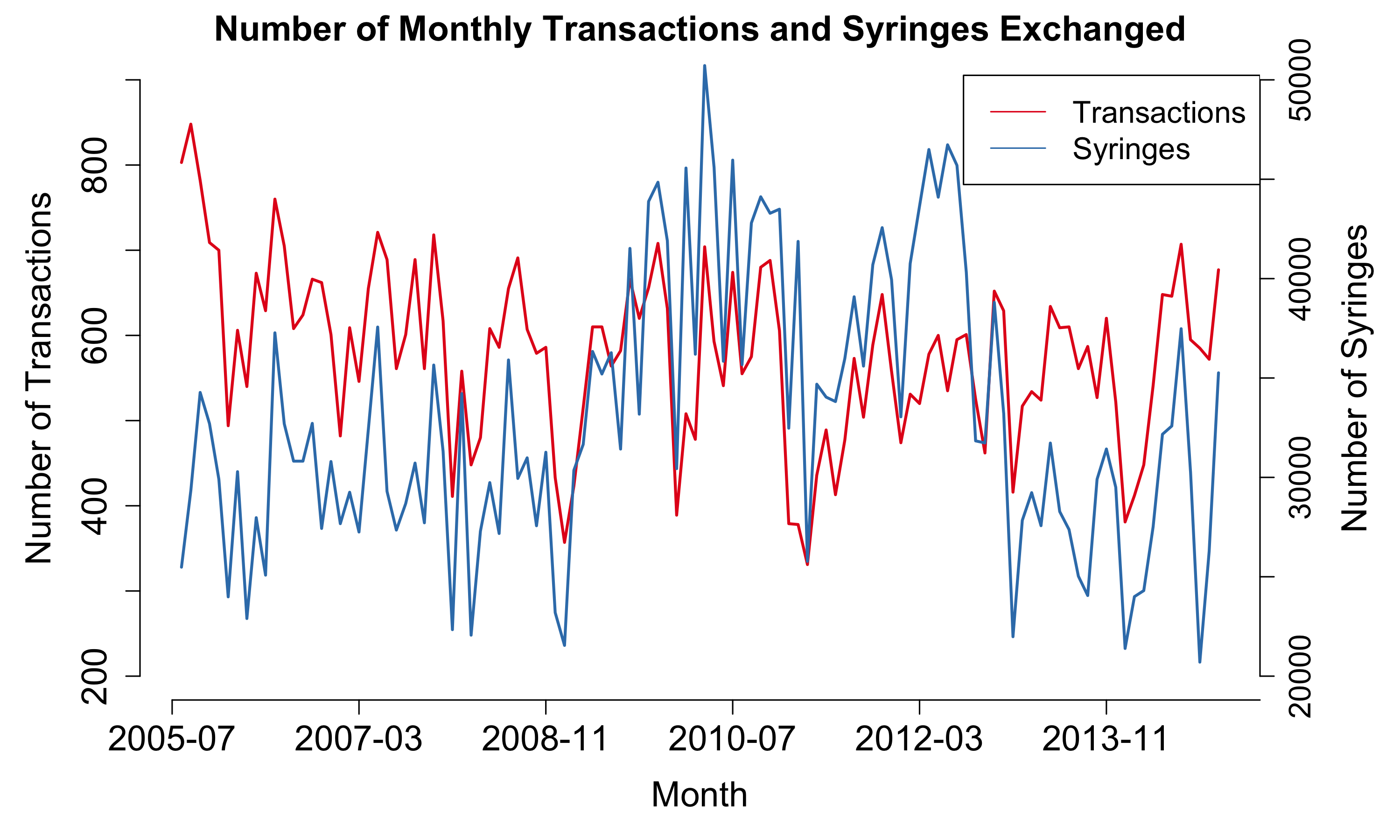}
		\caption{The time series of aggregated monthly transactions and syringes exchanged}
		\label{fig:Arrivals}
	\end{figure}

	We present a time series of the number of transactions from July 2005 to November 2014 in Figure~\ref{fig:TimeLoc}. The figure shows transactions at \tcb{four storefronts and mobile distribution in two} areas,  \textcolor{black}{where a mobile van was dispatched to certain locations with a flexible schedule. According to SEP staff, the schedule for the van was communicated to clients during their visits and at some shooting galleries. Clients could also inquire about the schedule through phone calls.} The service at Location \textcolor{black}{6 was} terminated in January 2010, and some of its clients \textcolor{black}{started} visiting Location 4 afterward, \tcb{consistent with} the increasing trend for Location 4. \tcb{Figure~\ref{fig:TimeLoc} also suggests a declining number of visits for Location 2.} Figure~\ref{fig:Arrivals} shows the fluctuation of monthly aggregated transactions. The figure suggests a slight decrease in the number of monthly transactions over the ten-year time span. The number of syringes exchanged \textcolor{black}{surged} between 2009 and 2012 but appeared to decrease afterward.
	
	\section{Model of Client Arrival Process} \label{sec:Model}
	We seek to develop a predictive model for a client's arrival process based on a number of predictors such as race, gender, age, and more. To this end, we segment a client's experience in the system into three sub-processes: initiation, reoccurring visits, and termination. Features of the client are used as covariates to estimate parameters of the model of reoccurring visits and termination. We can achieve two objectives with our model. First, we can forecast the next arrival of a specific client, given that client's features and most recent arrival time. Second, we can simulate the system and perform sensitivity analysis on specific model parameters. The former can help the SEP identify irregular behavior and take prompt intervention measures. The latter can guide initiatives to improve system-wide performance. 
	
	The overall structure of the model we formulate is that we build sub-models of these three individual sub-processes. While we have analytical models of these sub-components, our overall model, which combines these sub-models, can only be executed as a simulation, as we describe after characterizing the sub-models.  
	
	\subsection{Initiation} \label{subsec:initiation}
\begin{figure}
		\centering
		\includegraphics[width=\textwidth*3/4]{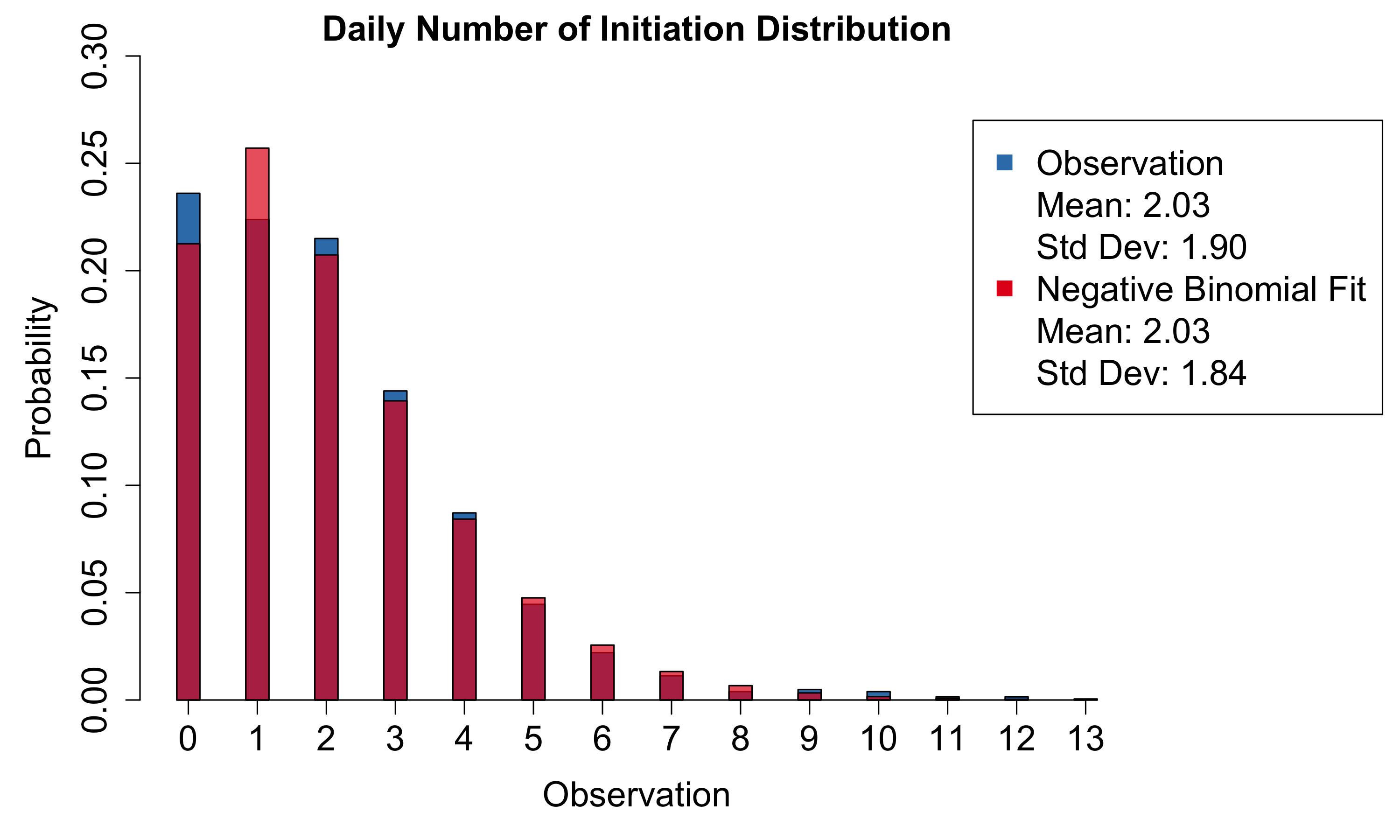}
		\caption{Empirical distribution fit of number of daily initiations to a negative binomial distribution}
		\label{fig:negBino}
	\end{figure}

	The first time that a client uses the SEP is called the \emph{initiation} of the client's arrival process. The recorded initiation data are clear, and so we focus on how to simulate new initiations. We examine the distribution of the number of initiations per day, i.e., the arrivals generated by clients who have never previously visited a service location. Since our SEP started recording survey data four years after recording transactions, some returning clients were asked to complete the survey starting in July 2005, even though their true initiation was earlier. In an attempt to avoid inflating some initiation counts, we use data starting from January 2007 in order to fit a distribution to estimate the initiation process. The blue bars in Figure~\ref{fig:negBino} show the empirical distribution of initiations per day. We fit a negative binomial distribution with parameter \((3, 0.59725)\), shown as the red bars in Figure~\ref{fig:negBino}. The negative binomial distribution can be seen as an over-dispersed version of a Poisson distribution and \tcb{is} used to model discrete data whose sample variance exceeds the sample mean; see, e.g., \cite{gardner1995regression}. In Section~\ref{subsec:Validation}, we detail goodness-of-fit measures for this and other distributional estimates. 
	
	For the purpose of simulation, each day we first generate a number of total new clients using the negative binomial distribution. In addition to simulating new arrivals of clients, we must assign attributes to those clients. In our simulation, we do so by drawing a client at random (with replacement) from the collection of 5,903 clients in our dataset. From the survey data, 33 numerical and categorical characteristics describe the client, and these are summarized in Appendix~\ref{appen:dataFit}. 
	
	\subsection{Reoccurring Visits} \label{subsec:reoccur}
\begin{figure}[H]
		\centering
		\includegraphics[width=\textwidth*3/4]{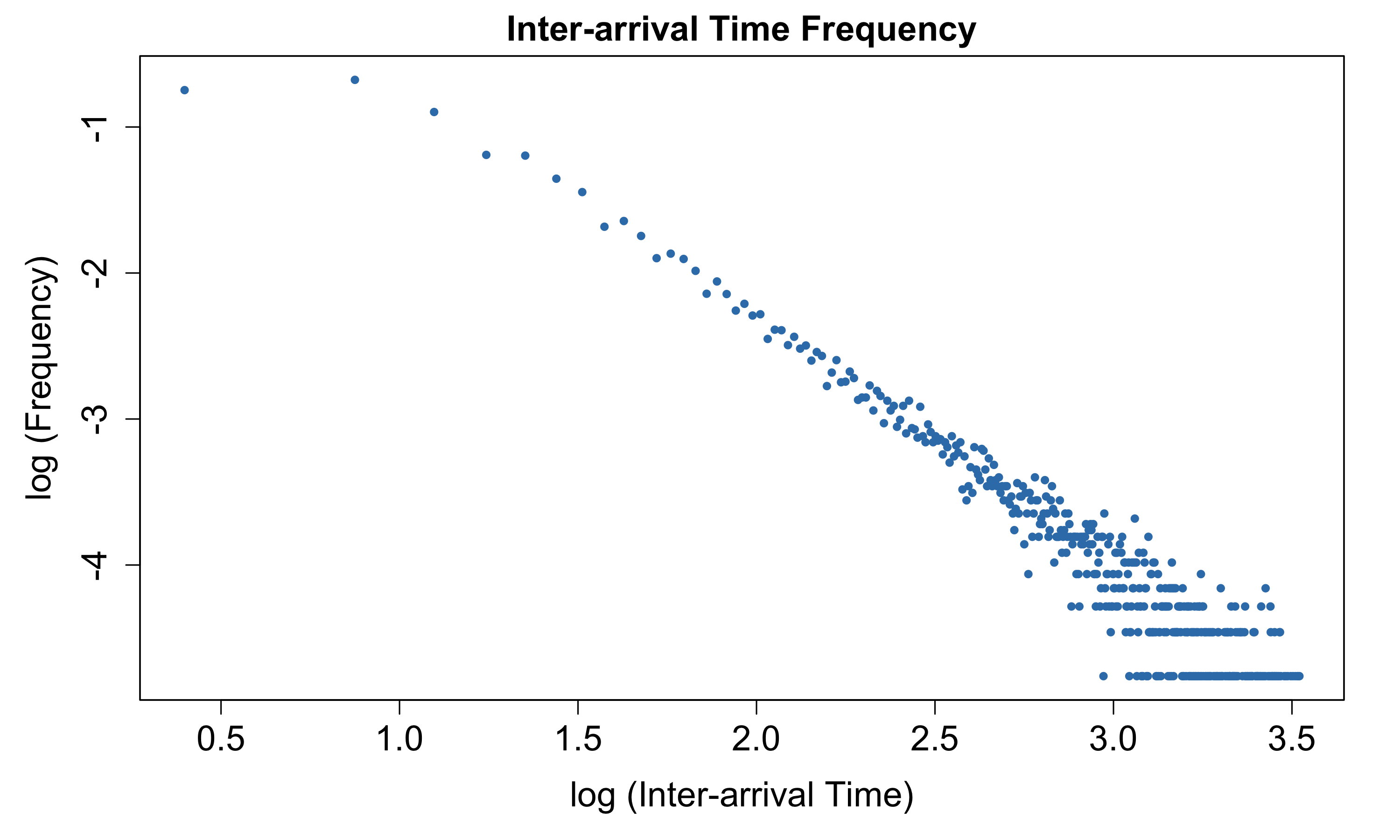}
		\caption{Log-log relationship between frequency and inter-arrival time. The logarithms in the figure are base 10, and underlying inter-arrival times are in days. }
		\label{fig:interArrLog}
	\end{figure}

	We track the history of clients who visit the SEP service sites multiple times and plot the distribution of inter-arrival times. The inter-arrival time is defined here as the duration between two consecutive visits made by the same client. Figure~\ref{fig:interArrLog} suggests that the distribution has a heavy tail, i.e., it shrinks to zero more slowly than an exponential. 
	
	\tcb{As we suggest in Section~\ref{sec:ProbDesc}, using a Poisson process to model inter-arrival times does not help with our main goal, which is to provide contextual, i.e.,  client-specific predictions.  Putting this aside for a moment, we tested the goodness of fit associated with a Poisson process for our aggregate dataset, and for datasets associated with sub-populations of clients based on ethnicity, age, and gender.  Moreover, we investigated both homogeneous and non-homogeneous (e.g., piecewise constant arrival rate by week) Poisson processes. For the models we assessed, statistical tests yielded \(p\)-values that were vanishingly small, suggesting that such models fail to provide an adequate representation. This is consistent with our observation from Figure~\ref{fig:interArrLog}, giving further evidence that modeling inter-arrival times using an exponential distribution may not be appropriate.} 
	
	We model client inter-arrival times with a phase-type distribution, because of its goodness of fit and its potential interpretability. The distribution of any nonnegative random variable can be approximated with high accuracy using a phase-type distribution; see, e.g., \cite{Asmussen2003Applied}. A phase-type distribution can be expressed as the time required for a continuous-time Markov chain (CTMC) to enter an absorbing state (say, state $0$) from a randomly selected transient state, $\{1,2,\ldots,n\}$; see, e.g., \cite{Buchholz14}. A probability mass function, denoted by \(\alpha=(\alpha_i)_{i =1,\ldots,n}\), governs the initial state of the CTMC. The infinitesimal generator is constructed in the following manner:
	\begin{equation}
	\label{eqn:qmatrix}
	\begin{bmatrix}
	0 & 0\\
	a & Q
	\end{bmatrix},
	\end{equation}
	where \(a\) is an \(n\)-dimensional vector specifying the transition rates from the transient states to the absorbing state, and \(Q\) is an 
	\(n \times n\) matrix specifying the transition rates among transient states, where \(Q(i,j)\) denotes the rate of transitioning from state \(i\) to state \(j\), and \(Q(i,i) = - [\sum_{j \neq i, j = 1, \dots, n} Q(i,j)+a(i)]\). The first row in the generator of equation~\eqref{eqn:qmatrix} corresponds to the transition rates from the absorbing state to any other transient state, which are always 0. The probability density function (pdf) of the phase-type distribution can be characterized as \(f(t) = \alpha e^{Qt}a\), and the cumulative distribution function (cdf) is given by \(F(t) = 1 - \alpha e^{Qt}\mathbf{1}\), where \(\mathbf{1}\) is the \(n\)-dimensional vector of all 1's.
	
	There are multiple ways to fit a phase-type distribution to data; see, e.g., \cite{nelson2010capturing}. Here, we formulate a nonlinear optimization model rooted in maximum-likelihood estimation (MLE), coupled with a regression model that uses covariates of the clients. We \tcb{first} describe the MLE approach in the context of the Coxian distribution, a special case of phase-type distributions, \tcb{and then couple it with client-specific predictors.} 
	
\begin{figure}[H]
		\centering
		\includegraphics[width=\textwidth*3/4]{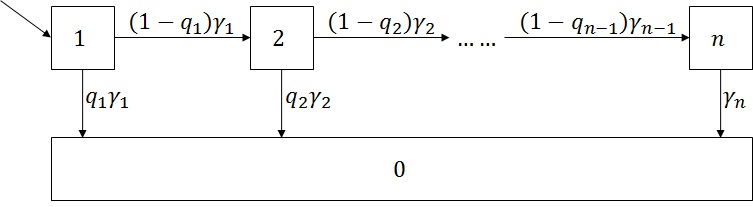}
		\caption{CTMC depiction of the Coxian distribution}
		\label{fig:Coxian}
	\end{figure}

	For a Coxian distribution, the embedded Markov chain has \(n + 1\) states, as shown in Figure~\ref{fig:Coxian}. The stochastic process starts in state 1, i.e., \(\alpha=(1,0,\ldots,0)\), and from each transient state, \(i=1,\ldots,n-1\), we can transition to only the adjacent transient state, \(i+1\) or to the absorbing state, \(0\). The rate at which we depart state \(i\) is \(\gamma_i\), and we transition to the absorbing state with probability \(q_i\) and to the adjacent transient state with probability \(1-q_i\). As Figure~\ref{fig:Coxian} also depicts, transient state \(n\) can only transition to the absorbing state.  \\
	\begin{figure}[H]
		\centering
		\includegraphics[width=\textwidth*3/4]{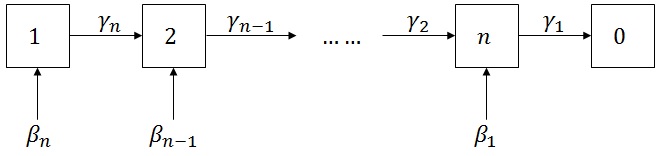}
		\caption{An equivalent CTMC to the Coxian distribution's model in Figure~\ref{fig:Coxian}}
		\label{fig:Coxian2}
	\end{figure}
	
	A Coxian distribution is more parsimonious than a general phase-type distribution. The former model contains \(2n-1\) parameters while the latter has up to \(n^2+n\). Bobbio and Cumani~\cite{Bobbio90} show that the Coxian model in Figure~\ref{fig:Coxian} is equivalent to another CTMC model that is depicted in Figure~\ref{fig:Coxian2}. The latter formulation is helpful in our setting because it offers a linear structure for capturing client-specific features as we describe below. The relationship between the \(q_i\) parameters in the first model and the \(\beta_i\) parameters in the second model is given by the following equations:
	\begin{align*}
	& q_0 = \beta_0 = 0&\\
	& \beta_i = q_i \prod_{k = 0}^{i-1} (1-q_k) & i = 1, 2, \dots, n\\
	& q_i = \frac{\beta_i}{1-\sum_{k = 0}^{i-1}\beta_k} & i = 1, 2, \dots, n.
	\end{align*} 
	
	A CTMC lends itself to interpretation by associating transitions in the model with assumed phases of a client using syringes after leaving a service location. \tcb{In addition to the references mentioned in Section~\ref{sec:ProbDesc}, which use CTMCs with hidden states \cite{chehrazi2019dynamic,paddock2012epidemiological,liu2015efficient}, we note that such an approach has also been used to model the length of stay of hospital patients \cite{Faddy09, Faddy99, Faddy05}, including work in which serial (Coxian) CTMCs are employed.} We use a similar philosophy by inferring transition rates from unobservable states and, moreover, connecting them to features of a client.
	
	We can interpret a Coxian distribution with \(n=2\) in our setting as follows. After visiting an SEP location, the client enters an (unobservable) ``active state'' with probability \(\beta_1\), and subsequently returns to an SEP site after an exponentially distributed delay with parameter \(\gamma_1\). Alternatively, with probability \(\beta_2=1-\beta_1\) the client enters a ``passive state.'' Returning to an SEP site is then the sum of two independent exponential random variables with rates \(\gamma_1\) and \(\gamma_2\), where we expect \(\gamma_1 > \gamma_2\). The passive state could correspond to the client temporarily seeking another source of syringes, for example. 
	
	Bobbio and Cumani \cite{Bobbio90} present an MLE procedure to fit the parameters for the Coxian distribution. Their method, however, is not directly applicable when we express \(\beta\) and \(\gamma\) as affine functions of predictors associated with clients. A result of \cite{Bibinger2013sum} allows us to express the pdf and cdf of a sum of independent exponential random variables, and we can use this result to write the pdf and cdf of a Coxian random variable as:
	\begin{equation}
	f(t) = \sum_{i = 1}^{n} \beta_{n+1-i} f_i(t) = \sum_{i = 1}^{n}\left[\beta_{n+1-i} \left(\prod_{l=1}^{n+1-i}\gamma_l\right)  \sum_{j=1}^{n+1-i}\frac{e^{-\gamma_j t}}{\prod_{k=1, k\neq j}^{n+1-i} \left(\gamma_k - \gamma_j\right)}\right] \label{eqn:fterm}
	\end{equation}
	\begin{equation}
	F(t) = \sum_{i = 1}^{n} \beta_{n+1-i} F_i(t) = \sum_{i = 1}^{n} \left[\beta_{n+1-i}  \left(\prod_{l=1}^{n+1-i}\gamma_l\right)  \sum_{j=1}^{n+1-i}\frac{ (1 - e^{-\gamma_j t})/\gamma_j}{\prod_{k=1, k\neq j}^{n+1-i} \left(\gamma_k - \gamma_j\right)}\right]. \label{eqn:Fterm}
	\end{equation}
	Here \(f_i\) and \(F_i\) are the pdf and cdf of the inter-arrival time, conditioned on beginning in state~\(i\). We index the collection of inter-arrival times by $\mathcal{S}$, and denote each inter-arrival time by $t_s$, $s \in \mathcal{S}$, and similarly denote censored inter-arrival times by $t_u$, $u \in \mathcal{U}$. For every client, the time from the last arrival to the end of the observation horizon can be considered a right-censored inter-arrival time. The likelihood function is the product of the pdf of each inter-arrival time and the complement of the cdf of each censored inter-arrival time. Maximizing the log-likelihood function then leads to the following problem:
	\begin{subequations}
		\begin{align}
		\max_{\beta,\gamma \geq 0} \quad & \sum_{s \in \mathcal{S}} \log(f(t_s))+ \sum_{u \in \mathcal{U}} \log \left(1 - F(t_u)\right) &\\
		\text{s.t.} \quad & \sum_{i = 1}^{n} \beta_i = 1 & \label{cons:sumBeta1} \\
		& f, F \text{ defined as in~\eqref{eqn:fterm} and~\eqref{eqn:Fterm}, } \forall t_s, s \in \mathcal{S} \text{ and } t_u, u \in \mathcal{U}, \text{ respectively}. \label{cons:fFterm}&
		\end{align}
		\label{prob:betagamma0}
	\end{subequations}
	The first term in the objective function of model~\eqref{prob:betagamma0} corresponds to observed inter-arrival times, and the second term corresponds to right-censored data in which we do not know the inter-arrival time, only that it exceeds, \(t_u\); see, e.g., \cite{Papaioannou2014Censor} for such treatments of right-censored data. We note that a limiting analysis shows that equations~\eqref{eqn:fterm}-\eqref{eqn:Fterm} remain valid even when rates at distinct states are identical \cite{Bibinger2013sum}. That said, this can cause numerical difficulties and we return to this issue below. 
	
	So far, the described fitting procedure assumes all clients behave according to the same model. As discussed above, we seek to incorporate the features of clients when we construct the parameters of the Coxian distribution. Here we use an affine relationship to connect those features to the parameters of the Coxian distribution. We use \(\mathcal{V}\) to represent the set of clients, and we use \(j = 1,2,\dots, m\) to index the characteristics of clients. We use \(x_{j,v},\ \forall j = 1, \dots, m,\ v \in \mathcal{V}\), to denote these predictors. We also use \(v(s) \in \mathcal{V}\) to specify the client associated with the \(s\)-th inter-arrival time, and we similarly define \(v(u)\) for the client associated with the \(u\)-th censored inter-arrival time. Our extension of model~\eqref{prob:betagamma0} to incorporate client-specific predictors is given by: 
	
	\begin{subequations}
		\small
		\begin{align}
		\max_{\beta,\gamma,b,g,\epsilon} \quad & \sum_{s \in \mathcal{S}} \log(f(t_s)) + \sum_{u \in \mathcal{U}} \log \left(1 - F(t_u) \right) - \eta^\beta \|\epsilon^\beta\|^2_2 \label{objFull}\\
		\text{s.t.} \quad & \sum_{i = 1}^{n} \beta_{i,v} = 1 \qquad \forall v \in \mathcal{V} \label{cons:sumBeta2}\\
		& \beta_{i,v} = \sum_{j = 1}^{m} b_{i,j}x_{j,v} + b_{i,0} + \epsilon^\beta_{i,v} \qquad \forall i = 1,2,,\dots,n, v \in \mathcal{V} \label{cons:linearFitbeta} \\
		& \gamma_{i,v} = \sum_{j = 1}^{m} g_{i,j}x_{j,v} + g_{i,0} \qquad \forall i = 1,2,\dots,n, v \in \mathcal{V} \label{cons:linearFitgamma}\\
		& f(t_s) = \sum_{i = 1}^{n}\left[\beta_{n+1-i,v(s)} \left(\prod_{l=1}^{n+1-i}\gamma_{l,v(s)}\right)  \sum_{j=1}^{n+1-i}\frac{e^{-\gamma_{j,v(s)} t_s}}{\prod_{k=1, k\neq j}^{n+1-i} \left(\gamma_{k,v(s)} - \gamma_{j,v(s)}\right)}\right] \;\; \qquad \forall s \in \mathcal{S} \label{cons:fterm}\\
		& F(t_u) = \sum_{i = 1}^{n} \left[ \beta_{n+1-i,v(u)} \left(\prod_{l=1}^{n+1-i}\gamma_{l,v(u)} \right)  \sum_{j=1}^{n+1-i}\frac{e^{-\gamma_{j,v(u)} t_u}/\gamma_{j,v(u)}}{\prod_{k=1, k\neq j}^{n+1-i} \left(\gamma_{k,v(u)} - \gamma_{j,v(u)}\right)} \right] \qquad \forall u \in \mathcal{U} \label{cons:Fterm}\\
		& \beta_{i,v} \geq 0 \qquad \forall i = 1,2,\dots, n, v \in \mathcal{V}\\
		& \gamma_{i,v} \geq 0 \qquad \forall i = 1,2,\dots, n, v \in \mathcal{V}.
		\end{align}
		\label{prob:betagamma}
	\end{subequations}
	The idea behind model~\eqref{prob:betagamma} is that we have predictors associated with each client, and constraints~\eqref{cons:linearFitbeta} and~\eqref{cons:linearFitgamma} express the parameters of the Coxian model as an affine function of these predictors. Constraint~\eqref{cons:sumBeta2} replicates constraint~\eqref{cons:sumBeta1}, and constraints~\eqref{cons:fterm} and~\eqref{cons:Fterm} define the pdf and cdf terms that appear in the log-likelihood in the first two terms of the objective function. 
	
	Model~\eqref{prob:betagamma} combines elements of regression and maximum likelihood estimation. The first two terms in the objective function maximize log-likelihood in the spirit of model~\eqref{prob:betagamma0}. Constraints~\eqref{cons:linearFitbeta}-\eqref{cons:linearFitgamma} define the regression model. Parameters \(\beta_{i,v}\) and \(\gamma_{i,v}\) are unobservable, and so, in principle, we could have no residual term in the regression model. However, constraint~\eqref{cons:sumBeta2} requires that the conditional probabilities that we return to states \(1,2,\dots, n\) sum to 1. So, to maintain feasibility we add a residual, \(\epsilon_v^\beta\), in equation~\eqref{cons:sumBeta2} and penalize its two-norm using a positive weight \(\eta^\beta\) in the final term in the objective function~\eqref{objFull}. While not explicit in model~\eqref{prob:betagamma}'s statement, to help prevent overfitting, we regularize the regression parameters \(b\) and \(g\) by adding terms \(-\eta^b \| b \|^2_2\) and \(-\eta^g \| g \|^2_2\) to the objective function~\eqref{objFull}.
	
	While a limiting analysis shows the validity of equation~\eqref{eqn:fterm}-\eqref{eqn:Fterm} even when some of the components of \(\gamma\) are identical, allowing this when optimizing can cause numerical problems. Moreover, our motivation, sketched above, includes the idea that the sojourn times should be larger in the passive state than in the active state, and for both reasons we add the following constraint to model~\eqref{prob:betagamma}: 
	\begin{equation}
	\label{gamma_gap}
	\gamma_{i,v} - \gamma_{i+1,v} \geq \delta, \ \ \forall i=1,2,\ldots,n-1, v \in \mathcal{V}, \text{ where } \delta > 0.
	\end{equation}
	
	\subsection{Termination} \label{subsec:termination}
	We assume that each client has a possibility to exit the system after each visit to the SEP. Specifically, as soon as the CTMC hits the absorbing state, we assume that, with probability \(p_v\), client~\(v \in \mathcal{V}\) will stop visiting our SEP. 
	
	However, we cannot observe a client leaving the system because we only observe their visits. If a client visits service locations multiple times, then we know that for every visit before the last one, the client is still in the system, and so the likelihood function is conditioned on the client remaining in the system. After the last visit, a client may stay in the system or may leave. We need to incorporate this information in the likelihood function. Conditioned on the client remaining in the system, the likelihood function is \(1 - F(t_u)\), where \(t_u,\ u \in \mathcal{U}\), is the time between the client's last visit and the end of the observation horizon. As a result, the log-likelihood function can be revised as:
	\begin{equation}\label{cons:pfFterm}
	\sum_{s \in \mathcal{S}} \left(\log(f(t_s)) + \log (1 - p_{v(s)}) \right) + \sum_{u \in \mathcal{U}} \log \left(\left(1 - F(t_u)\right) (1 - p_{v(u)}) + p_{v(u)} \right).
	\end{equation}
	We again model parameter \(p_v\) via a functional relationship with client \(v\)'s covariates. Instead of an affine relationship, we use a logistic \tcb{function as follows}:
	\begin{equation} \label{cons:pFit}
	p_v = \left(1 + e^{-(\rho_0 + \sum_{j = 1}^m \rho_j x_{j,v})} \right)^{-1},
	\end{equation}
	where we will optimize the fit via parameters \(\rho_{j}\) and \(\rho_0\). Given that \(p_v \in (0,1)\), the logistic function is a natural choice; we do note that we also tested a linear relationship but obtained poorer results.
	
	We can fit the termination parameter \(p\) by combining the results of~\eqref{prob:betagamma},~\eqref{cons:pfFterm}, and~\eqref{cons:pFit}. Since the likelihood function is conditioned on whether the client has exited the system, we need to solve a nonlinear optimization problem as follows, which fits parameters for both reoccurring visits and termination, \tcb{integrating our latter two sub-models:}
	
	\begin{subequations}
		\label{prob:betagammap}
		\small
		\begin{align}
		\max_{\beta,\gamma,b,g,\rho,p,\epsilon} \quad & \sum_{s \in \mathcal{S}} \left(\log(f(t_s)) + \log (1 - p_{v(s)}) \right) + \sum_{u \in \mathcal{U}} \log \left(\left(1 - F(t_u)\right) (1 - p_{v(u)}) + p_{v(u)} \right) - \eta^\beta \|\epsilon^\beta\|^2_2 \label{objFullP}\\
		\text{s.t.} \quad & \sum_{i = 1}^{n} \beta_{i,v} = 1 \qquad \forall v \in \mathcal{V} \label{cons:sumBetap}\\
		& \beta_{i,v} = \sum_{j = 1}^{m} b_{i,j}x_{j,v} + b_{i,0} + \epsilon^\beta_{i,v} \qquad \forall i = 1,2,,\dots,n, v \in \mathcal{V} \label{cons:linearFitbetap} \\
		& \gamma_{i,v} = \sum_{j = 1}^{m} g_{i,j}x_{j,v} + g_{i,0} \qquad \forall i = 1,2,\dots,n, v \in \mathcal{V} \label{cons:linearFitgammap}\\
		& p_v = \left(1 + e^{-(\rho_0 + \sum_{j = 1}^m \rho_j x_{j,v})} \right)^{-1} \qquad \forall v \in \mathcal{V} \label{cons:linearFitpp}\\
		& f(t_s) = \sum_{i = 1}^{n}\left[\beta_{n+1-i,v(s)} \left(\prod_{l=1}^{n+1-i}\gamma_{l,v(s)}\right)  \sum_{j=1}^{n+1-i}\frac{e^{-\gamma_{j,v(s)} t_s}}{\prod_{k=1, k\neq j}^{n+1-i} \left(\gamma_{k,v(s)} - \gamma_{j,v(s)}\right)}\right] \;\;\qquad \forall s \in \mathcal{S} \label{cons:ftermp}\\
		& F(t_u) = \sum_{i = 1}^{n} \left[ \beta_{n+1-i,v(u)} \left(\prod_{l=1}^{n+1-i}\gamma_{l,v(u)} \right)  \sum_{j=1}^{n+1-i}\frac{e^{-\gamma_{j,v(u)} t_u}/\gamma_{j,v(u)}}{\prod_{k=1, k\neq j}^{n+1-i} \left(\gamma_{k,v(u)} - \gamma_{j,v(u)}\right)} \right] \qquad \forall u \in \mathcal{U} \label{cons:Ftermp}\\
		& \beta_{i,v} \geq 0 \qquad \forall i = 1,2,\dots, n, v \in \mathcal{V}\\
		& \gamma_{i,v} \geq 0 \qquad \forall i = 1,2,\dots, n, v \in \mathcal{V}.
		\end{align}
	\end{subequations}
	
	Solving problem~\eqref{prob:betagammap} maximizes the log-likelihood function for the combination of reoccurring visits and termination. Similar to fitting the reoccurring visits, we also add a regularization term \(-\eta^\rho\|\rho\|^2_2\) in the objective function to prevent overfitting. Constraint~\eqref{cons:linearFitpp} models the logistic relationship between the parameter \(p\) and the covariates \(x\). Given the fit value of \(\rho^*\) and the features of client \(v\), we can calculate the termination probability of that client,  \(p_v\).
	
	\section{Experimental Results} \label{sec:experiment}
	In this section, we first discuss how we solve model~\eqref{prob:betagammap} and its simpler variants, along with preliminary results in which we do \emph{not} use the covariates of the clients.  Then we present the results of the fit model, provide insights as to how different features predict client behavior, test elements of model validity, and show an example of how our personalized arrival model can guide active intervention.
	
	\subsection{Computational Issues and Preliminary Results} \label{subsec:prelim}
	We use a Coxian model with \(n=2\) transient states. We interpret one phase as the client being in an active state with the SEP, i.e., with more frequent visits to exchange syringes, and we interpret the other phase as a passive state with less frequent visits.
	
	Model~\eqref{prob:betagammap} and its variants are computationally challenging nonconvex optimization problems. We use Ipopt 3.12.1 \cite{wachter2006implementation}, with linear solver MA27, to solve instances of these optimization problems. Due to nonconvexity, we only obtain locally optimal solutions. In addition, because numerical issues can arise, we briefly sketch ways in which we ``help'' the solver. 
	
	We scale all continuous data, i.e., client predictor data, so that it is normalized. As discussed at the end of Section~\ref{subsec:reoccur}, we enforce \(\gamma_{1,v} \geq \gamma_{2,v}+\delta\), and we use \(\delta=0.005\) in our computation. For numerical reasons, we also bound the \(\gamma\)-parameters away from zero, by enforcing \(\gamma_2 \geq \underline{\gamma} \equiv 0.0005\).
	
	We start by solving a simplified variant of model~\eqref{prob:betagammap} in which we remove the predictors and directly optimize \(\beta, \gamma\), and \(p\). We do this for two reasons. First, it provides insight regarding typical values of these parameters, and second, as we discuss in further detail below, it helps provide a good initial solution for model~\eqref{prob:betagammap}. In particular we solve:
	\begin{subequations}
		\label{prob:betagammap0}
		\begin{align}
		\max_{\beta,\gamma,p} \quad & \sum_{s \in \mathcal{S}} \left(\log(f(t_s)) + \log(1-p)\right) + \sum_{u \in \mathcal{U}} \log \left((1 - F(t_u))(1-p) + p \right) \\
		\text{s.t.} \quad & \sum_{i = 1}^{n} \beta_{i} = 1 \\
		& f(t_s) = \sum_{i = 1}^{n}\left[\beta_{n+1-i} \left(\prod_{l=1}^{n+1-i}\gamma_{l}\right)  \sum_{j=1}^{n+1-i}\frac{e^{-\gamma_{j} t_s}}{\prod_{k=1, k\neq j}^{n+1-i} \left(\gamma_{k} - \gamma_{j}\right)}\right] \qquad \forall s \in \mathcal{S} \\
		& F(t_u) = \sum_{i = 1}^{n} \left[ \beta_{n+1-i} \left(\prod_{l=1}^{n+1-i}\gamma_{l} \right)  \sum_{j=1}^{n+1-i}\frac{e^{-\gamma_{j} t_u}/\gamma_{j}}{\prod_{k=1, k\neq j}^{n+1-i} \left(\gamma_{k} - \gamma_{j}\right)} \right] \qquad \forall u \in \mathcal{U} \\
		& \gamma_{i} - \gamma_{i+1} \geq \delta \qquad \forall i = 1, 2, \dots, n-1\\
		& \beta_{i} \geq 0 \qquad \forall i = 1,2,\dots, n\\
		& \gamma_{n} \geq \underline{\gamma}\\
		& 0 \leq p \leq 1.
		\end{align}
	\end{subequations}
	
	Solving model~\eqref{prob:betagammap0} leads to parameters for a ``featureless'' client, as follows: 
	\[\hat{\beta}_1 = 0.8194,\ \hat{\beta}_2 = 0.1806\]
	\[\hat{\gamma}_1 = 0.0520,\ \hat{\gamma}_2 = 0.0030 \]
	\[\hat{p} = 0.0981.\]
	
	This result suggests that after each visit, the featureless client has a \(9.8\%\) chance of exiting the SEP system. Conditional on the client visiting an SEP site again, the client returns via the active state (frequent visits) with a probability of about \(0.82\). The mean \tcb{return time to the SEP} from this state is \(1/\gamma_1 \approx 19\) days. With probability about \(0.18\), the client returns via the passive state, and the expected time to visit the SEP is then \(1/\gamma_2 + 1/\gamma_1 \approx 350\) days. 
	
	Rather than optimizing over the intercept terms, \(b_0\), \(g_0\) and \(\rho_0\) in model~\eqref{prob:betagammap}, we fixed these terms as:
	\[b_{1,0} = \hat{\beta}_1 \quad b_{2,0} = \hat{\beta}_2\]
	\[g_{1,0} = \hat{\gamma}_1 \quad g_{2,0} = \hat{\gamma}_2\]
	\[\rho_0 = \ln\left(\frac{\hat{p}}{1 - \hat{p}}\right).\]
	Fixing the intercept terms in this way allows us to interpret parameters \(b_{i,j}\), \(g_{i,j}\), and \(\rho_j\) for \(j=1,2,\ldots,m\) as deviations from the featureless client. Moreover, fixing these parameters helps improve the numerical performance of Ipopt when solving the nonconvex problem, in part by effectively providing a good initial solution. 
	
	After adding the regularization terms for \(b\), \(g\) and \(\rho\) described in Section~\ref{subsec:reoccur} and~\ref{subsec:termination}, and fixing the value of \(b_0\), \(g_0\) and \(\rho_0\), we solve the following nonlinear program:
	
	\begin{subequations}
		\small
		\begin{align}
		\max_{\beta,\gamma,b,g,\rho,p,\epsilon} \quad & \sum_{s \in \mathcal{S}} \left(\log(f(t_s)) + \log (1 - p_{v(s)}) \right) + \sum_{u \in \mathcal{U}} \log \left(\left(1 - F(t_u)\right) (1 - p_{v(u)}) + p_{v(u)} \right) \nonumber\\
		& - \eta^\beta \|\epsilon^\beta\|^2_2 - \eta^b \|b\|_2^2 - \eta^g \|g\|_2^2 - \eta^\rho \|\rho\|_2^2\\
		\text{s.t.} \quad & \sum_{i = 1}^{n} \beta_{i,v} = 1 \qquad \forall v \in \mathcal{V} \label{cons:sumBeta2N}\\
		& \beta_{i,v} = \sum_{j = 1}^{m} b_{i,j}x_{j,v} + b_{i,0} + \epsilon^\beta_{i,v} \qquad \forall i = 1,2,,\dots,n, v \in \mathcal{V} \label{cons:linearFitbetaN} \\
		& \gamma_{i,v} = \sum_{j = 1}^{m} g_{i,j}x_{j,v} + g_{i,0} \qquad \forall i = 1,2,\dots,n, v \in \mathcal{V} \label{cons:linearFitgammaN}\\
		& p_v = \left( 1 + e^{-(\rho_0 + \sum_{j = 1}^m \rho_j x_{j,v})} \right)^{-1} \qquad \forall v \in \mathcal{V} \label{cons:linearFitpN}\\
		& f(t_s) = \sum_{i = 1}^{n}\left[\beta_{n+1-i,v(s)} \left(\prod_{l=1}^{n+1-i}\gamma_{l,v(s)}\right)  \sum_{j=1}^{n+1-i}\frac{e^{-\gamma_{j,v(s)} t}}{\prod_{k=1, k\neq j}^{n+1-i} \left(\gamma_{k,v(s)} - \gamma_{j,v(s)}\right)}\right] \;\;\quad \forall s \in \mathcal{S} \label{cons:ftermN}\\
		& F(t_u) = \sum_{i = 1}^{n} \left[ \beta_{n+1-i,v(u)} \left(\prod_{l=1}^{n+1-i}\gamma_{l,v(u)} \right)  \sum_{j=1}^{n+1-i}\frac{e^{-\gamma_{j,v(u)} t_u}/\gamma_{j,v(u)}}{\prod_{k=1, k\neq j}^{n+1-i} \left(\gamma_{k,v(u)} - \gamma_{j,v(u)}\right)} \right] \quad \forall u \in \mathcal{U} \label{cons:FtermN}\\
		& \gamma_{i,v} - \gamma_{i+1,v} \geq \delta \qquad \forall i = 1, 2, \dots, n-1, v \in \mathcal{V} \label{cons:gammaDiff2N}\\
		& \beta_{i,v} \geq 0 \qquad \forall i = 1,2,\dots, n, v \in \mathcal{V}\\
		& \gamma_{n,v} \geq \underline{\gamma} \qquad \forall v \in \mathcal{V}\\
		& b_{i,0} = \hat{\beta}_i, g_{i,0} = \hat{\gamma}_i, \rho_0=\ln\left( \frac{\hat{p}}{1 - \hat{p}} \right) \qquad \forall i=1,2,\ldots,n.
		\end{align}
		\label{prob:betagammaN}
	\end{subequations}
	With modest tuning effort, we select the following weights on the regularization terms:
	\[\eta^\beta = 100,\ \eta^b = 100,\ \eta^g = 1000,\ \eta^\rho = 10. \] 
	
	\subsection{Results and Analysis}\label{subsec:fitting}
	The results from fitting the parameters using the method in Section~\ref{subsec:prelim} are displayed in Table~\ref{tab:fitted1}. The estimators are represented by $\rho$,  \(b\), and \(g\). A positive \(\rho_j, j = 1,2,\dots, m\), indicates that having feature \(j\) increases the probability that the client will exit the system. Given that the client stays in the system, a positive value of parameter \(b_{1,j}\) increases the probability that the client returns to the active state. We do not report \(b_{2,j}\) because its coefficient differs from \(b_{1,j}\)'s by a sign. Positive coefficients \(g_{1,j}\) and \(g_{2,j}\) lead to increased frequencies, i.e., shorter mean times, associated with the active and passive states, respectively. The coefficients in Table~\ref{tab:fitted1} are given in either regular font or \textcolor{lightgray}{gray font}. The former category is significant, and the latter is not, where ``significant'' is defined as having at least 90\% of bootstrapped replications having the same sign, as detailed in Appendix~\ref{appen:coeffSig}.
	
	\tcb{Column \(\Delta\) in Table~\ref{tab:fitted1} shows the amount by which the conditional expected inter-arrival time changes (in days) if a client has that row's feature but is otherwise a featureless client. Column~\(T\) similarly shows the magnitude by which the expected sojourn time in the system changes due to a single feature.}
	For context, the mean sojourn time of a featureless client is \(806.5\) days. The acronyms FUD, FROS, and FBSA in the table respectively stand for frequency of using drugs, frequency of reusing own syringes, and frequency of being the area of an SEP location.
	\begin{table}[h]
		\centering
		\begin{tabular}{ l | l | r | r | r | r | r | r |}
			\hline
			\(\qquad\)& Factor (\(j\))	&	\(\rho_j\)	&	\(b_{1,j}\)	&	\(g_{1,j}\)	&	\(g_{2,j}\)	&	\(\Delta\)	&	\(T\)	\\ \hline
			& Gallery	&	0.5448	&	-0.0387	&	\textcolor{lightgray}{0.0027}	&	\textcolor{lightgray}{0.0003}	&	\textcolor{lightgray}{5.9}	&	-268.2	\\ \hline
			& From Other Locations	&	0.2047	&	\textcolor{lightgray}{-0.0156}	&	\textcolor{lightgray}{0.0023}	&	0.0004	&	\textcolor{lightgray}{-3.2}	&	-162.2	\\ \cline{2-8}
			&From Other SEP	&	-0.3156	&	0.0248	&	-0.0063	&	\textcolor{lightgray}{0.0000}	&	\textcolor{lightgray}{-6.2}	&	186.3	\\ \cline{2-8}
			& From Friends	&	-0.1194	&	\textcolor{lightgray}{-0.0045}	&	\textcolor{lightgray}{-0.0011}	&	-0.0007	&	20.9	&	330.1	\\ \cline{2-8}
			& From Strangers	&	-0.2697	&	\textcolor{lightgray}{0.0142}	&	\textcolor{lightgray}{0.0020}	&	0.0010&	-19.3	&	\textcolor{lightgray}{-26.8}	\\ \hline
			& Speedball	&	\textcolor{lightgray}{-0.0293}	&	-0.0704	&	0.0146	& -0.0008	&	48.1	&	524.9	\\ \cline{2-8}
			& Heroin	&	-0.0437	&	\textcolor{lightgray}{-0.0258}	&	\textcolor{lightgray}{0.0061}	&	-0.0008	&	31.4	&	365.6	\\ \hline
			& In Treatment	&	-0.3254	&	0.0243	&	-0.0126	&	\textcolor{lightgray}{0.0002}	&	\textcolor{lightgray}{-4.7}	&	215.7	\\ \cline{2-8}
			& Been in Treatment	&	-0.1137	&	-0.0178	&	0.0063	&	\textcolor{lightgray}{-0.0001}	&	\textcolor{lightgray}{5.9}	&	154.8	\\ \hline
			& Female	&	\textcolor{lightgray}{0.0154}	&	\textcolor{lightgray}{0.0217}	&	\textcolor{lightgray}{-0.0076}	&	0.0008	&	-14.8	&	-159.8	\\ \hline
			& White	&	\textcolor{lightgray}{0.0000}	&	0.0218&	\textcolor{lightgray}{-0.0005}	&	\textcolor{lightgray}{0.0002}	&	-9.9&	-101.0\\ \cline{2-8}
			& African American	&	0.3623&	\textcolor{lightgray}{-0.0073}	&	\textcolor{lightgray}{-0.0026}	&	\textcolor{lightgray}{0.0001}	&	\textcolor{lightgray}{1.6}	&	-209.6\\ \cline{2-8}
			& Puerto Rican	&	-0.7594	&	\textcolor{lightgray}{0.0084}	&	0.0101&	\textcolor{lightgray}{0.0001}	&	\textcolor{lightgray}{-7.4} &	673.3\\ \cline{2-8}
			& Mexican	&	-0.3383&	\textcolor{lightgray}{0.0014}	&	\textcolor{lightgray}{0.0014}&	0.0010&	-15.6&	\textcolor{lightgray}{76.0}	\\ \cline{2-8}
			& Other	&	-0.1994&	\textcolor{lightgray}{0.0088}	&	\textcolor{lightgray}{-0.0054}	&	\textcolor{lightgray}{0.0000}	&	\textcolor{lightgray}{-0.7}	&	151.9\\ \hline
			& Age of First Drug Use	&	\textcolor{lightgray}{1.1219}	&	0.0117&	\textcolor{lightgray}{0.0007}	&	0.0003&	-8.8&	\textcolor{lightgray}{-525.7}	\\ \cline{2-8}
			& Drug Use Span	&	\textcolor{lightgray}{1.8506}	&	-0.0042& -0.0030	&	\textcolor{lightgray}{-0.0001}	&	4.4&	\textcolor{lightgray}{-602.3}	\\ \cline{2-8}
			& FUD	&	\textcolor{lightgray}{0.0332}	&	0.0073&	\textcolor{lightgray}{-0.0011}	&	-0.0001	&	\textcolor{lightgray}{0.4}	&	\textcolor{lightgray}{-19.6}	\\ \cline{2-8}
			& FROS	&	0.0380&	\textcolor{lightgray}{-0.0051}	&	\textcolor{lightgray}{0.0000}	&	\textcolor{lightgray}{0.0000}	&	\textcolor{lightgray}{1.7}	&	\textcolor{lightgray}{-10.5}	\\ \cline{2-8}
			& FBSA	&	-0.1968	&	\textcolor{lightgray}{-0.0033} & 0.0058	&	0.0002	&	-5.2&	95.4\\ \hline
			\multicolumn{1}{r}{\tiny{Note:}}	&\multicolumn{7}{l}{\tiny{Here, $\rho_j$, $b_{1,j}$, and $g_{1,j}$/$g_{2,j}$ are factor-specific regression coefficients for the probability of exiting the system,}} \\ [-1.25ex]
			\multicolumn{1}{l}{ \ }	&\multicolumn{7}{l}{\tiny{probability of returning to the active state, and mean transition times in the CTMC, respectively. Based on the}} \\ [-1.25ex]
			\multicolumn{1}{l}{ \ }	&\multicolumn{7}{l}{\tiny{factor in each row, parameter $\Delta$ denotes the amount by which the conditional expected inter-arrival time changes,}} \\ [-1.25ex]
			\multicolumn{1}{l}{ \ }	&\multicolumn{7}{l}{\tiny{and  $T$ similarly denotes changes in the system sojourn time, both in days.}} \\
		\end{tabular}
		\caption{Fitted parameters of the Coxian process}\label{tab:fitted1}
	\end{table}		
	
	Table~\ref{tab:fitted1} provides information on how a client's attributes affect the probability the client leaves the system and the frequency with which the client makes use of SEP services, even though the factors are from different categories, e.g., type of drugs the client uses versus where the client obtains syringes versus ethnicity. Some observations from the table include: 
	\begin{enumerate}
		\item If the client attends a shooting gallery, it is more likely for the client to exit the system or become passive. The former factor dominates in that the overall expected time in the system decreases relative to a featureless client (\(T\)). 
		\item Clients who can obtain syringes from other locations, such as pharmacies, are more likely to exit the system, perhaps because they are not reliant on SEP services. On the other hand, if the client obtains syringes from other sources (other SEPs, friends, and strangers), which may not be as reliable, it is more likely for the client to remain in the system.
		\item The \(b_1\)-coefficient associated with speedball (a type of drug mixing cocaine with heroin or morphine) is strongly negative, meaning the client is less likely to stay in the active state, leading to an increase in expected inter-arrival time. 
		\item A client in a treatment program is more likely to stay in the SEP system, as indicated by a negative value of \(\rho\) and a positive value of \(T\).
		\item A female client is more likely to visit frequently, but remains in the system for a shorter period of time.
		\item The probability of exiting the system differs significantly according to ethnicity: African-American clients are more likely to exit the system while the opposite is true for Puerto Rican and Mexican clients. 
		\item Not surprisingly, clients who are more frequently near an SEP site (i.e., have larger values of FBSA) are less likely to leave the system, are more likely to visit a site frequently, and overall have longer sojourn times in the system. 
	\end{enumerate}
	
	Such observations from the model we fit may allow our SEP to tailor promotion of their services to specific target populations. For example, the starkly different behavior of African-American clients may warrant special attention from the SEP in early encounters, relative to PWIDs of other ethnicities. Bean~\cite{bean2016cocaine} states that African-American drug users are less likely to inject drugs than White drug users. However, among all clients, it is not clear why African Americans are less likely to seek the syringe exchange services offered by our SEP. Investigating whether African Americans are more likely to quickly abandon injection drug use, versus continue use but not seek SEP services, would likely be needed to guide such strategies. Our results show that it would be beneficial to increase the frequency of clients being in the service location area, and one possible solution is to increase the frequency of the mobile van service in certain locations to improve accessibility. 
	
	\subsection{Model Validation} \label{subsec:Validation}
	The model of Section~\ref{sec:Model} quantifies client behavior. In this section, we perform statistical tests and compare results of the simulation outputs with observed data in order to assess model validity. 
	
	\tcb{We begin with further details on the implementation of the simulation model. We simulate the arrival of clients to the SEP over 2,310 days, equivalent to the number of days that our SEP was open between July 2005 and November 2014. We also simulate an initial 5,000 days as a warm-up period, since our SEP was established more than 15 years before 2005. We assume SEP staff are always available to serve a visiting client, there are no shortages of syringes or other resources that would alter client behavior, and the location and operating schedule of storefronts and mobile vans are fixed. In other words, we assume the nature of client visits is governed solely by the features of the clients, in a manner consistent with historical data, and is not affected by service or resource availability. This matches our understanding of the actual system, as we discuss in Section~\ref{sec:Data}. In our simulation of 7,310 days in total, for each day} we first simulate the number of new clients according to the negative binomial distribution. We assign features to these clients by drawing a client at random, with replacement, from the list of 5,903 clients described in Section~\ref{sec:Data}. Given the features of a client, \(x\), and the parameters, \(\rho\), \(b\), and \(g\), \textcolor{black}{obtained from model~\eqref{prob:betagammaN}, we can calculate the parameters of the CTMC, \(\beta_i, \gamma_i,\ i = 1,2\), and \(p\), by:
		\begin{subequations}
			\begin{align}
			&\beta_i = \sum_{j = 1}^{m} b_{i,j}x_{j} \qquad i = 1,2 \label{eqn:beta_affine}   \\
			&\gamma_i = \sum_{j = 1}^{m} g_{i,j}x_{j} \qquad i = 1,2 \label{eqn:gamma_affine}  \\
			&p = \left( 1 + e^{-(\rho_0 + \sum_{j = 1}^m \rho_j x_{j})} \right)^{-1}. \label{eqn:depart_prob}
			\end{align}
		\end{subequations}
		With the CTMC built for every client, upon the arrival of a specific client, we simulate whether this client exits the system using $p$ from equation~\eqref{eqn:depart_prob}. And, if the client does not exit the system, we simulate the time of the next arrival using the $\beta$ and $\gamma$ values from equations~\eqref{eqn:beta_affine}-\eqref{eqn:gamma_affine}. After the warm-up period and simulating the 2,310 days of interest, we obtain summary statistics as outputs based on inter-arrival times, sojourn times, and number of visits of each client.}
	
	We perform statistical tests to assess the quality of our sub-models for initiation, reoccurring visits, and termination. For initiation, we selected the negative binomial distribution for reasons that we discuss in Section~\ref{subsec:initiation}. Using a Pearson's chi-squared goodness-of-fit test we obtain a $p$-value of 0.250, suggesting that we should not reject the null hypothesis that the data are consistent with the fit distribution. For comparison, we also fit other commonly used distributions (geometric, binomial, uniform, Poisson, and hyper-geometric), and we performed the same goodness-of-fit tests. None of those distributions had a $p$-value that exceeded a 0.05 level of significance.
	
\begin{figure}
		\captionsetup[subfloat]{font=footnotesize}
		\subfloat[][Coxian Inter-arrival Model]{\label{fig:interArrComparison_ll_a}\includegraphics[width=0.48\textwidth]{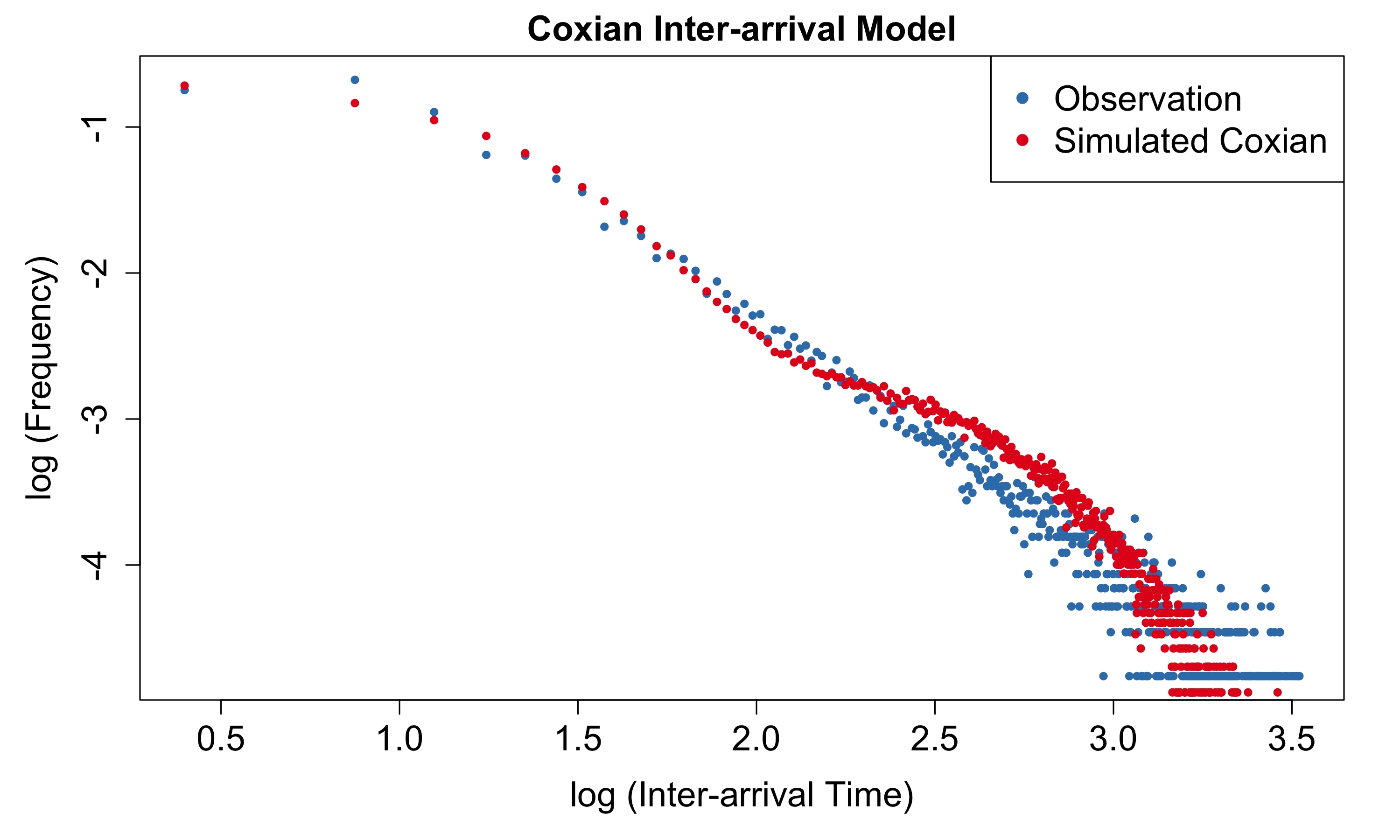}}       
		\hspace*{0.10in}        
		\subfloat[][Exponential Inter-arrival Model]{\label{fig:interArrComparison_ll_b}\includegraphics[width=0.49\textwidth]{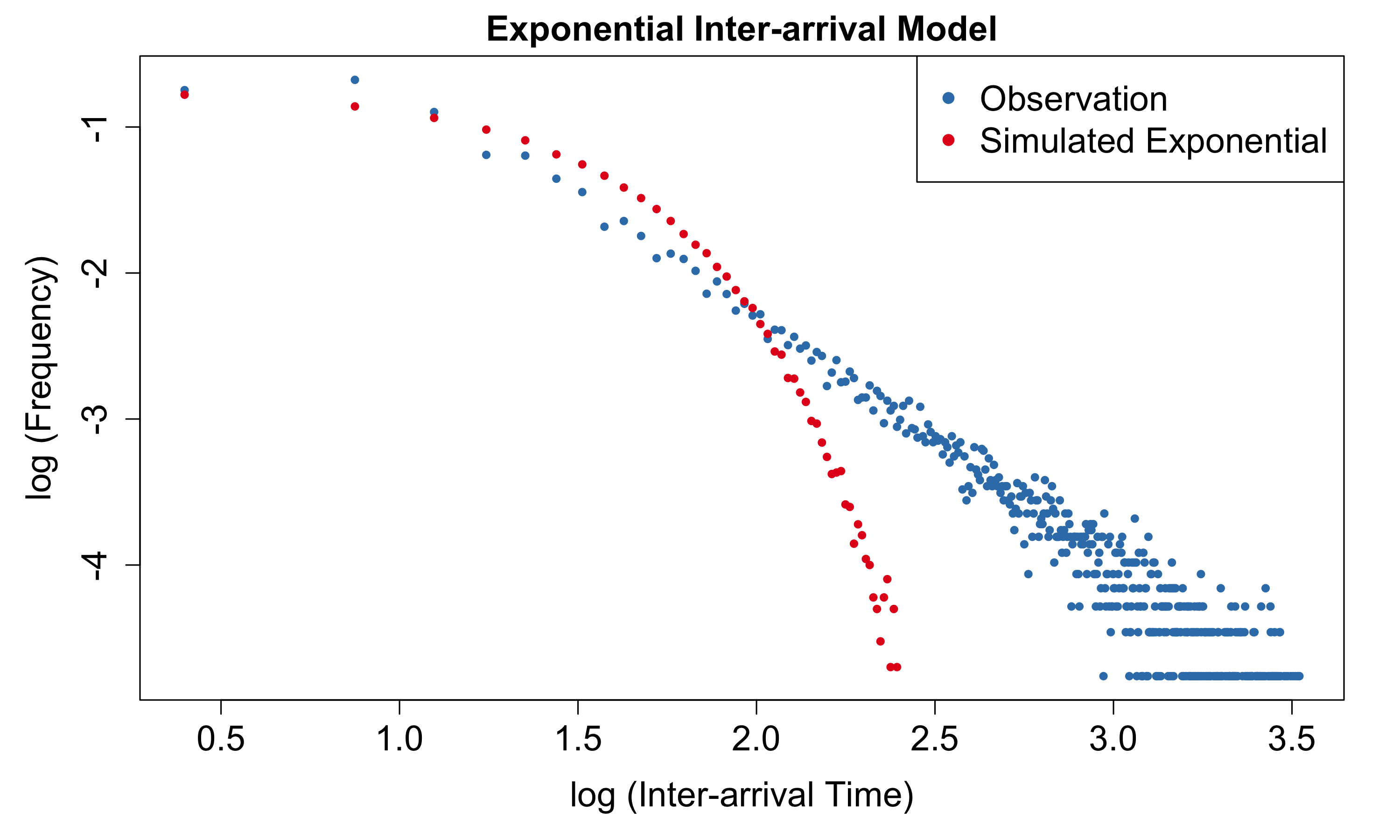}} 
		\caption{\tcb{Part~(a) of the figure shows the log-log relationship between frequency and the Coxian inter-arrival time model. Part~(b) shows the analogous relationship  for the exponential inter-arrival time model. In both subplots the (simulated) model values are shown with red dots and observed data are shown with blue dots.} The  logarithms are base 10 and the underlying inter-arrival times in days.
		}\label{fig:interArrComparison_ll}
	\end{figure}

	For reoccurring visits, we compare the distribution of inter-arrival times obtained from the simulation model with an exponential distribution with a mean of \(67.50\) days, which is the mean of observed inter-arrival times. Figure~\ref{fig:interArrComparison_ll} illustrates such a comparison. Results from the simulated Coxian process are shown on the left, and those from the exponential distribution are on the right, both in red. In addition, we plot actual observations in blue. The figure suggests that our Coxian-based simulation model provides a better fit to the observed data.
	
\begin{figure}
		\centering
		\includegraphics[width=0.75\textwidth]{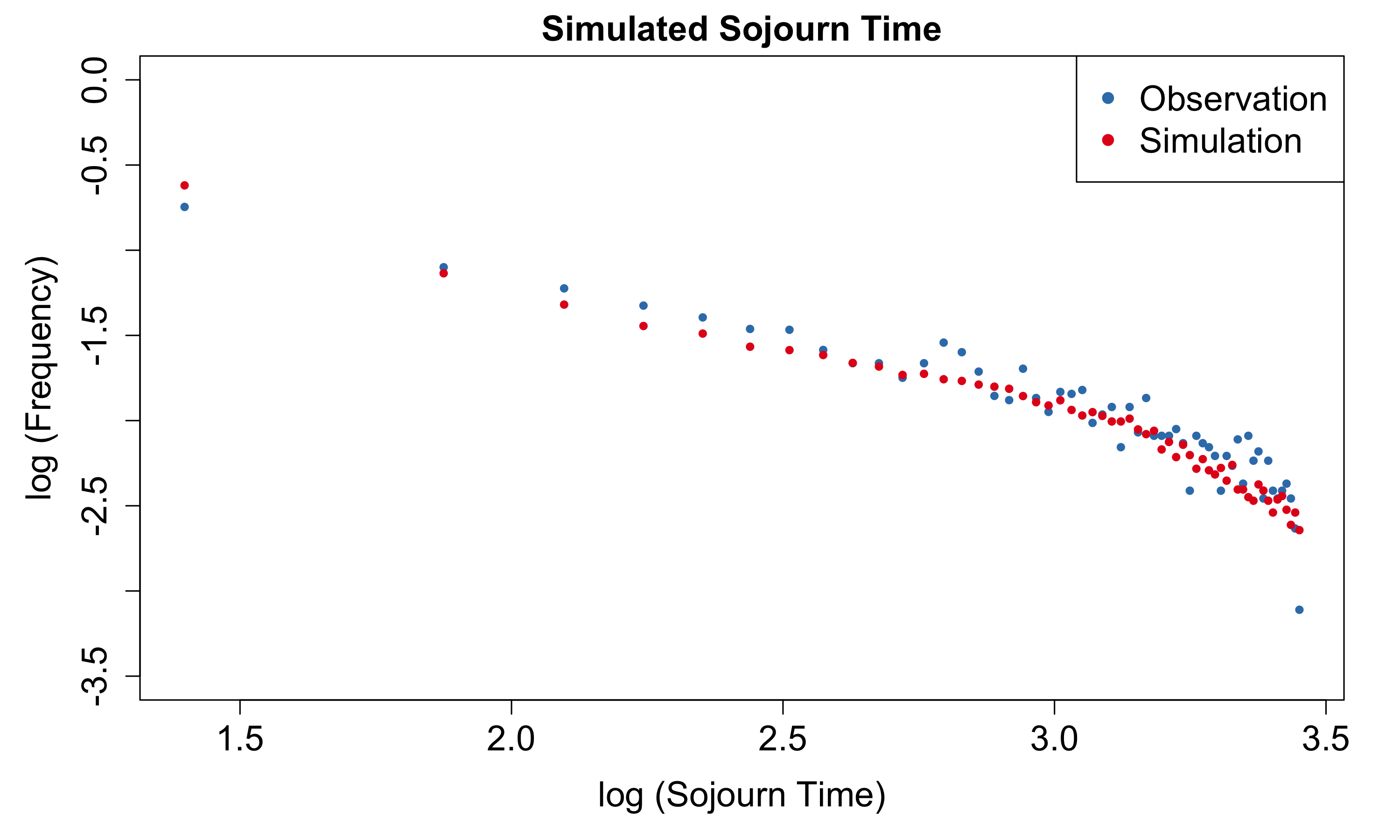}
		\caption{The log-log (base 10) relationship between frequency and simulated sojourn time is shown with red dots, and the same relationship between frequency and observed sojourn time is shown with blue dots.}
		\label{fig:sojournComp}
	\end{figure}

	Since we do not directly observe whether a client has left the system, we test the right-censored sojourn time of clients in the system. We run a Pearson's chi-squared procedure to test the null hypothesis that the simulated distribution is consistent with observed data. The \(p\)-value of the test is \(0.626\), which suggests that we should not reject the null hypothesis. Figure~\ref{fig:sojournComp} suggests that the simulated sojourn times from the simulation appear consistent with observed data.
	
	In addition to assessing the validity of our three sub-models, Table~\ref{table:conditionalMean} suggests that our simulation model can accurately capture demographic information such as ethnicity. The second and the third columns represent new clients by ethnicity. These columns are, of course, very close because we simply draw from the observed set of clients. The values help give a sense of variability due to sampling. The two right-most columns are the percentage of visits to SEP sites based on ethnicity. The consistency of these simulated values with observed percentages hinges on the Coxian model accurately capturing ethnicity-based inter-arrival times and exits from the system. For example, the simulation model captures well the lower values for African Americans, and larger values for Puerto Ricans, relative to their rates of initiation in the system.
	\begin{table}[H]
		\centering
		\begin{tabular}{| l | l | l | l | l |}
			\hline
			\multirow{2}{*}{Ethnic Group} & \multicolumn{2}{c |}{Initiation} & \multicolumn{2}{ c |}{SEP Locations Visits}  \\ \cline{2-5}
			& Observed & Simulated & Observed & Simulated\\
			\hline
			White& 51.58\% & 51.80\% & 47.53\% & 48.05\%\\
			\hline
			African American & 23.45\% & 23.66\% & 16.36\% & 15.99\%\\
			\hline
			Puerto Rican & 14.78\% & 14.21\% & 24.16\% & 23.62\%\\
			\hline
			Mexican & 6.17\% & 6.23\% & 7.64\% & 8.00\%\\
			\hline
			Other Latino & 1.22\% & 1.17\% & 1.14\% & 1.17\%\\
			\hline
			Other & 2.80\% & 2.93\% & 3.17\% & 3.16\%\\
			\hline
		\end{tabular}
		\caption{\tcb{Comparison of the observed and simulated percentages of clients for each ethnic group, and the observed and simulated percentage of arrivals for each ethnic group}}
		\label{table:conditionalMean}
	\end{table}
	
	\subsection{Guiding Active Intervention}\label{subsec:example}
	
	\textcolor{black}{Our simulation model can facilitate analysis to provide SEP staff with insights regarding: (i)~specific clients who are likely to enter a passive state or exit the system, and (ii)~dispatch policies for the mobile van. We discuss both of these in turn.} 
	
	\subsubsection{Simple Client-Specific Intervention}
	While not immediate from Table~\ref{tab:fitted1}, our discrete-event simulation model can be used to estimate that a 40-year old Puerto Rican male, with a history of using drugs for 21 years, who injects heroin 10 times per day, wants treatment, uses syringes after others once every 30 days, and frequents the area of an SEP site, has a 95\% chance of having entered the passive state if he does not visit a service location within 56 days, \textcolor{black}{assuming he has not already terminated contact with the SEP system}. As a result, SEP staff could send a text message to such a client as a reminder if he has not returned within two months. 
	
	\subsubsection{Intervention with Mobile Van Dispatch}
	\textcolor{black}{The value of actively reaching out to clients based on insights from our simulation model may be further enhanced by {\em mobile} exchange of syringes. Here, we simulate mobile van dispatch, with a personalized notification push, to show its potential to improve current SEP operations. The corresponding simulation model has the following constructs and assumptions. }
	
	\tcb{
		{\bf Clients.} We simulate the initiation and reoccurring visits of clients in the same way described in Section~\ref{subsec:Validation} with the following exception: We assume that when a client exits the system, there is a probability, denoted by \(p_r\), that the client is eligible to return to the system with active intervention. We say that a client is {\em at risk} if they reuse syringes, either their own or the syringes of another, and we note that 22.4\% of our 5,903 clients are at risk from the survey data.}
	
	\tcb{
		{\bf Mobile van.} We assume the SEP has one van, and each weekday the van is dispatched to one of five ZIP codes. We further assume that any client that the SEP contacts within a five-mile radius of that ZIP code is eligible to be served by the van. We selected the five ZIP codes by solving a facility-location model, which maximizes coverage of at-risk clients. The van visits each of these five ZIP codes in turn, Monday-Friday, each week over the simulation horizon.}
	
	\tcb{
		{\bf Risky behavior.} We assume only at-risk clients engage in risky behavior. And, we assume an at-risk client does {\em not} engage in risky behavior if the client is in the CTMC's active state, but otherwise the client does exhibit risky behavior. }
	
	\tcb{{\bf Intervention.} The SEP cannot observe a client's state or behavior, and hence intervention decisions are made knowing the time since the client's last visit and the client's predictors. In particular, $\gamma_1$ is the rate associated with the exponential distribution governing the return time, {\em if} the client is in the active state. We assume the SEP contacts a client if: (i)~the time since the last visit exceeds the 0.9-level quantile for the active state's exponential return-time distribution, and (ii)~the client's ZIP code is within five miles of the ZIP code the van is visiting that day.}
	
	\tcb{{\bf Re-engaging a client.} An intervention can be successful if the client is in the active state, passive state, or has exited but is eligible to return (with probability $p_r$). Among these clients, we let $p_s$ denote the probability that a contacted client will visit the van, and hence re-engage with the SEP. If a client ignores the notification, we assume the client stays in the same state: active, passive, or exited. We further assume the SEP stops contacting a client after {\em three} notification attempts have been ignored.}
	
\begin{figure}
		\centering
		\includegraphics[width=0.75\textwidth]{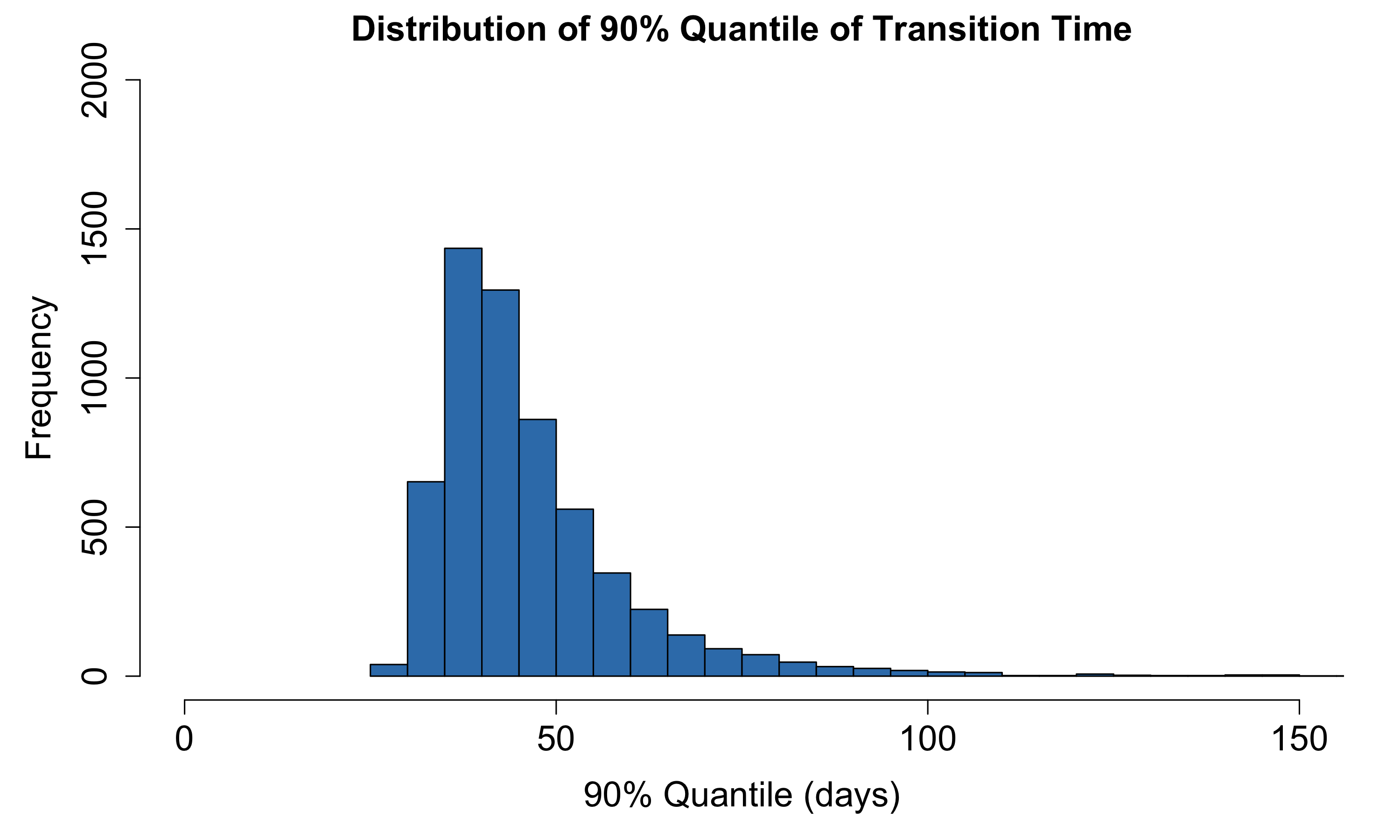}
		\caption{Histogram of the 90\% client-specific quantile values for the return-time distributions from the CTMC's active state for the 5,903 surveyed clients.}
		\label{fig:quantile}
	\end{figure}

	\textcolor{black}{Using the simulation model, we compare active intervention with current SEP operations, and we estimate the relative effectiveness by examining: (i) the number of additional arrivals to the system, and (ii) the number of times an SEP intervention re-engages a client who would otherwise be engaging in risky behavior. We estimate $p_r$ as follows. The 0.975-level quantile of the observed inter-arrival time is 552 days, and we use this as a proxy for whether the client has exited the system. We denote the number of inter-arrival times exceeding 552 days by \(N_r\), and the number of clients who have not visited any service location after 552 days by \(N_e\). We estimate \(p_r\) via:
		\begin{equation*}
		p_r = \frac{N_r}{N_r + N_e}.
		\end{equation*}
		With our data, \(N_r = 1354\) and \(N_e = 4281\), and so \(p_r \approx 0.24\).
		As we indicate above, we use the 0.9-level quantile for client-specific return-time distributions from the active state to make intervention decisions, and Figure~\ref{fig:quantile} shows a histogram of these quantiles among the surveyed clients. 
	}
	
	\textcolor{black}{We run the simulation model under two different system designs: (i) with active intervention via van dispatch and client notification, and (ii) without active intervention, approximating current SEP operations. The results are shown in Table~\ref{table:vanSimu} as we parametrically vary the success probability,~\(p_s\). After the warm-up period of 5,000 days, the average number of daily arrivals in the last 2,310 days has a mean of 24.6 and a 95\% confidence interval halfwidth of 0.9. For each value of \(p_s\) in the table, we run the simulation 20 times and present 95\% confidence intervals. The ``Total'' column under ``No.\ of Interventions" shows the number of times over 2,310 days that a notified client visited the mobile van, and the ``Risky'' column indicates the subset of those successful interventions involving at-risk clients in the passive or exited state.}
	\begin{table}[H]
		\centering
		\begin{tabular}{| l | c | c | c |}
			\hline
			\multirow{2}{*}{\(p_s\)} & Average Daily & \multicolumn{2}{c |}{No. of Interventions} \\ \cline{3-4}
			& Arrivals & Risky & Total \\ \hline
			0 & \(24.6 \pm 0.9\) & \(0\) & \(0\) \\ \hline
			0.01 & \(27.1 \pm 0.9\) & \(79.1 \pm 16.4 \) & \(398.3 \pm 37.3\) \\ \hline
			0.03 & \(28.2 \pm 1.0\) & \(254.5 \pm 38.4\) & \(1251.6 \pm 86.2\) \\ \hline
			0.05 & \(29.7 \pm 0.9\) & \(439.1 \pm 54.0\) & \(2171.9 \pm 117.7\) \\ \hline
			0.07 & \(31.0 \pm 1.1\) & \(641.5 \pm 66.4\) & \(3173.0 \pm 162.4\) \\ \hline
			0.09 & \(32.4 \pm 1.1\) & \(852.3 \pm 95.9\) & \(4235.3 \pm 217.8\) \\ \hline
		\end{tabular}
		\caption{Simulation results of nominal SEP operations ($p_s=0$) and SEP with active intervention via van dispatch and client notification. Here, $p_s$ denotes the success probability associated with client notification, and the $\pm$ values represent 95\% confidence interval halfwidths.}
		\label{table:vanSimu}
	\end{table}
	\textcolor{black}{Under current SEP operations ($p_s=0)$, we see about 56,800 client arrivals over the 2,310-day horizon. 
		Table~\ref{table:vanSimu} shows that a success probability of \(p_s=0.05\) leads to about one intervention per day over that time horizon, i.e., less than 4\% of the nominal total number of arrivals would be the immediate result of an intervention. However, because a successful intervention leads to a re-engaged client returning to the active state, we see a significant 20\% growth of about five arrivals per day relative to the $p_s=0$ case. And, this corresponds to about one client per week who stopped risky behavior because of the intervention. Having only one client per day visiting the van may seem low,  but a regularly scheduled van would also attract other clients not included in our model. More importantly, a 20\% growth in average daily arrivals would be a welcome improvement countering some of the trends we point to in Section~\ref{subsec:transaction}. We see these results as suggesting that there is value in the SEP investigating an active intervention scheme similar to our example.  Moreover, our simulation model makes it possible to exploit the contextual nature of our model of inter-arrival times to determine a personalized threshold for push notification.}
	
	\section{Conclusions} \label{sec:Conclusions}
	In this paper, we examine the survey and transaction data of one major syringe exchange service provider in the Chicago metropolitan area. We find there is a discrepancy between a slightly decreasing trend in the number of client transactions with our SEP and an increasing number of heroin users in Chicago and the {United States}. We also discover significant differences in the behavior of clients in terms of how they engage with the SEP based on demographic attributes and further personal characteristics. Based on our observations, we focus on producing personalized predictions for clients that can aid the SEP in improving the system such as intervention initiatives for clients with certain attributes.
	
	Standard stochastic models, such as Poisson processes, fail to accurately capture the observed inter-arrival process. Therefore, we formulate a CTMC-based simulation model to represent a client's path through the system. Our model consists of three sub-models: initiation, reoccurring visits, and termination, with their parameters learned from linear and logistic regression models integrated into the procedure by which we estimate the model's parameters. With the aid of this model, SEP staff and researchers can analyze the system parameters to draw useful conclusions for groups with different traits, so that proper actions can be taken towards a specific target group, or even the individual PWID. The quantitative model, combined with the personal interaction with each client can inform SEP staff of timely intervention opportunities. Such personalized recommendations may be particularly useful when the SEP faces challenges in tracking a large number of clients. Our simulation model can also help SEP staff evaluate the effectiveness of candidate initiatives. 
	
	\textcolor{black}{In the future, our method could be enhanced by finding an algorithm to fit the parameters for higher fidelity Markov chain models since our optimization model for parameter estimation depends on the closed-form representation of the Coxian distribution. Other functional forms for the predictive models could be integrated into the fitting procedure as well. Further sensitivity analysis could be performed to evaluate other potential initiatives beyond dispatching the mobile van. Optimization over the location and route of the van could also be investigated using our simulation platform.}
	
	\section*{Acknowledgments}
	We thank Dr.~Lawrence J. Ouellet from the University of Illinois at Chicago, Dr.~Alexander Gutfraind from Loyola University, Chicago, Dan Bigg from CRA and many other colleagues from COIP and CRA who provided insights and access to data that greatly assisted the research. \tcb{We are grateful to two referees and an Associate Editor, whose suggestions significantly improved the paper.}
	
	\bibliographystyle{iise_new}
	\bibliography{COIP_Paper_V8}
	\pagebreak
	\appendix
	\section{Covariates for the PWIDs Population}\label{appen:dataFit}
	When a client initiation occurs in the simulation, we draw a client at random, with replacement, from our dataset of 5,903 unique clients. In this appendix we detail those attributes, \tcb{beginning with the fraction of the population with each categorical attribute:} \smallskip
	\begin{table}[H]
		\centering
		\begin{tabular}{| c | c | c |}
			\hline 
			\multicolumn{3}{|c|}{Gender} \\ \hline
			Male & Female & Transgender\\
			\hline
			0.6947 & 0.3049 & 0.0004\\
			\hline
		\end{tabular}
	\end{table}
	\vspace*{-0.2in}
	\begin{table}[H]
		\centering
		\begin{tabular}{| c | c | c | c | c | c |}
			\hline
			\multicolumn{6}{|c|}{Ethnicity} \\ \hline
			White & African American  & Puerto Rican & Mexican & Other Latino & Other \\
			\hline
			0.5158 & 0.2345 & 0.1478 & 0.0617 & 0.0122 & 0.0280\\
			\hline
		\end{tabular}
	\end{table}
	\vspace*{-0.2in}
	\begin{table}[H]
		\centering
		\begin{tabular}{| c | c |}
			\hline
			\multicolumn{2}{|c|}{Snort before injection} \\ \hline			
			Yes & No \\
			\hline
			0.3446 & 0.6554\\
			\hline
		\end{tabular}
	\end{table}
	\vspace*{-0.2in}
	\begin{table}[H]
		\centering
		\begin{tabular}{| c | c |}
			\hline
			\multicolumn{2}{|c|}{Participation in} \\ 
			\multicolumn{2}{|c|}{shooting galleries} \\
			\hline	
			Yes & No \\
			\hline
			0.0864 & 0.9136\\
			\hline
		\end{tabular}
	\end{table}
	\vspace*{-0.2in}
	\begin{table}[H]
		\centering
		\begin{tabular}{| c | c | c | c | c |}
			\hline
			\multicolumn{5}{|c|}{Participation in treatment programs} \\
			\hline
			& Currently in & Been in & Tried to get into & Interested in\\
			\hline
			Yes & 0.1011 & 0.1870 & 0.0923 & 0.4738\\
			\hline
			No & 0.8989 & 0.8130 & 0.9077 & 0.5362\\
			\hline
		\end{tabular}
	\end{table}
	\vspace*{-0.2in}
	\begin{table}[H]
		\centering
		\begin{tabular}{| c| c| c| c| c| c|}
			\hline
			\multicolumn{6}{|c|}{Drugs used in the past 30 days} \\
			\hline
			& Speedball & Heroin & Cocaine & Ritalin Heroin & Other\\
			\hline
			Yes & 0.0486 & 0.9582 & 0.0581 & 0.0005 & 0.0185\\
			\hline
			No & 0.9514 & 0.0418 & 0.9419 & 0.9995 & 0.9815\\
			\hline
		\end{tabular}
	\end{table}
	\vspace*{-0.2in}
	\begin{table}[H]
		\centering
		\begin{tabular}{| c| c| c| c| c| c| c|}
			\hline
			\multicolumn{7}{|c|}{Source of syringes} \\
			\hline
			& Family & Friends & Acquaintance & Strangers & Other SEP & Other\\
			\hline
			Yes & 0.0586 & 0.2529 & 0.0530 & 0.0163 & 0.1360 & 0.6131\\
			\hline
			No & 0.9414 & 0.7471 & 0.9470 & 0.9837 & 0.8640 & 0.3869\\
			\hline
		\end{tabular}
	\end{table}
	\vspace*{-0.2in}
	\begin{table}[H]
		\centering
		\begin{tabular}{| c | c |}
			\hline
			\multicolumn{2}{|c|}{Reuse own syringes} \\
			\hline
			Yes & No \\
			\hline
			0.1579 & 0.8521\\
			\hline
		\end{tabular}
	\end{table}
	\vspace*{-0.2in}
	\begin{table}[H]
		\centering
		\begin{tabular}{| c | c |}
			\hline
			\multicolumn{2}{|c|}{Use syringes} \\
			\multicolumn{2}{|c|}{behind others} \\
			\hline
			Yes & No \\
			\hline
			0.1972 & 0.8028\\
			\hline
		\end{tabular}
	\end{table}
	\noindent For each of the continuous factors, we present descriptive statistics and histograms of their distributions. We use \(\mu\) to denote the mean of the factor and \(\sigma\) to denote the standard deviation.
	\begin{itemize}
		\setlength\itemsep{0.1em}
		\item Age: \(\mu = 34.79, \ \sigma = 11.22 \);
		\item Age of clients at their first injection: \(\mu = 23.44, \ \sigma = 7.86 \);
		\begin{figure}[H]
			\centering
			\includegraphics[width=0.8\textwidth]{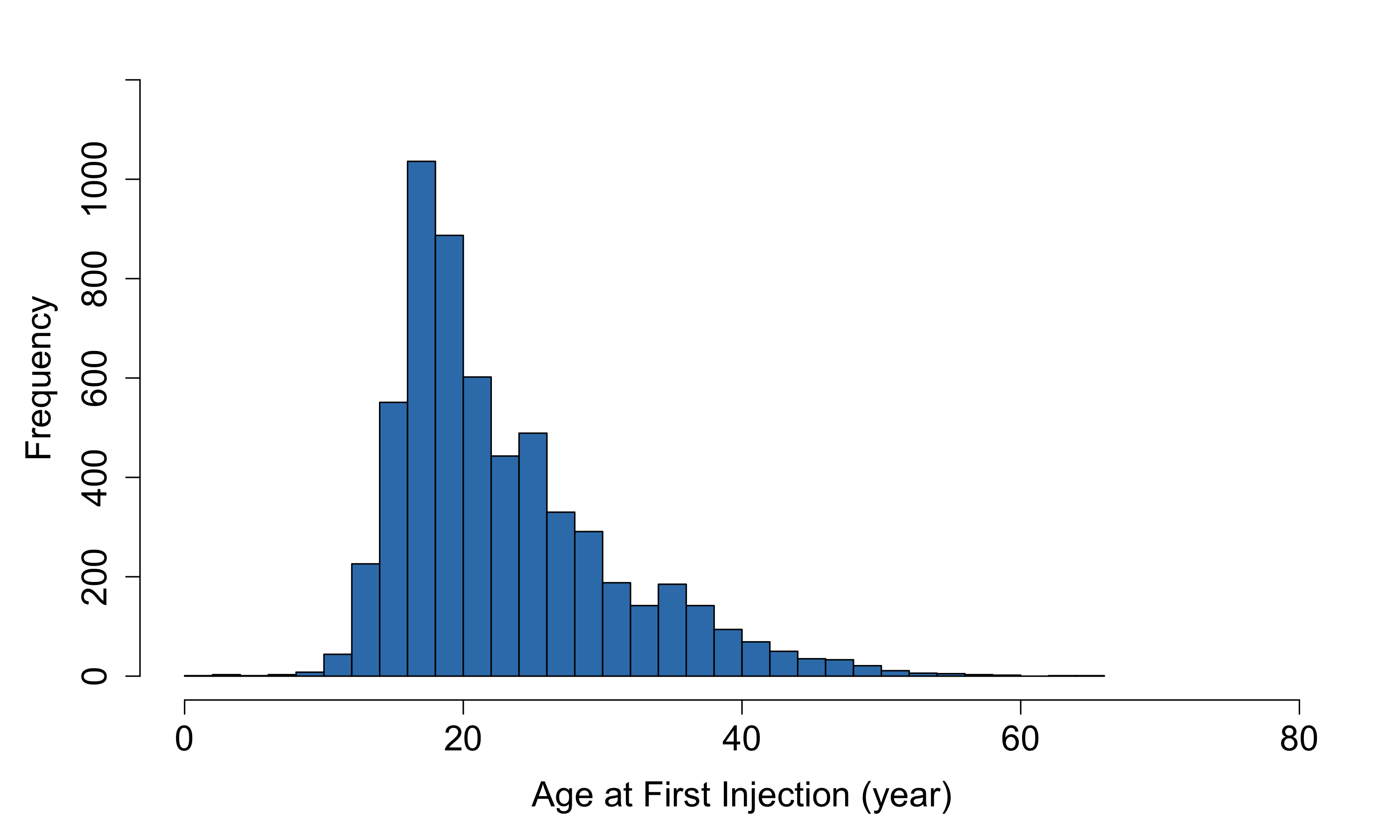}
			\caption{Distribution of the age of clients at their first injection}
			\label{fig:ageFirstDistr}
		\end{figure}
		\item Length of drug injection history: \(\mu = 11.36, \ \sigma = 11.36\);
		\begin{figure}[H]
			\centering
			\includegraphics[width=0.8\textwidth]{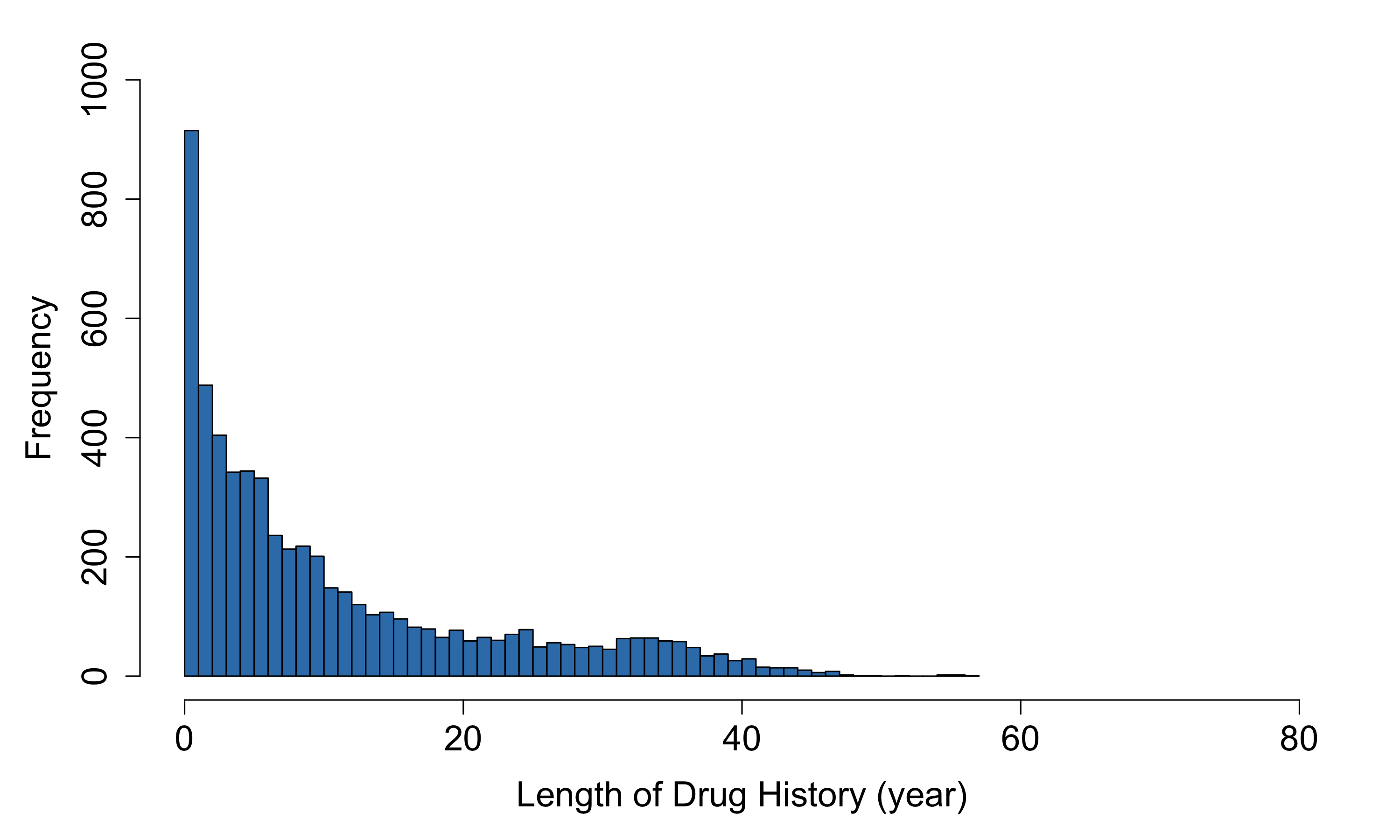}
			\caption{Distribution of the length of drug injection history}
			\label{fig:drugSpanDistr}
		\end{figure}
		\item Number of daily drug injections: \(\mu = 2.77,\ \sigma = 1.87\);
		\begin{figure}[H]
			\centering
			\includegraphics[width=0.8\textwidth]{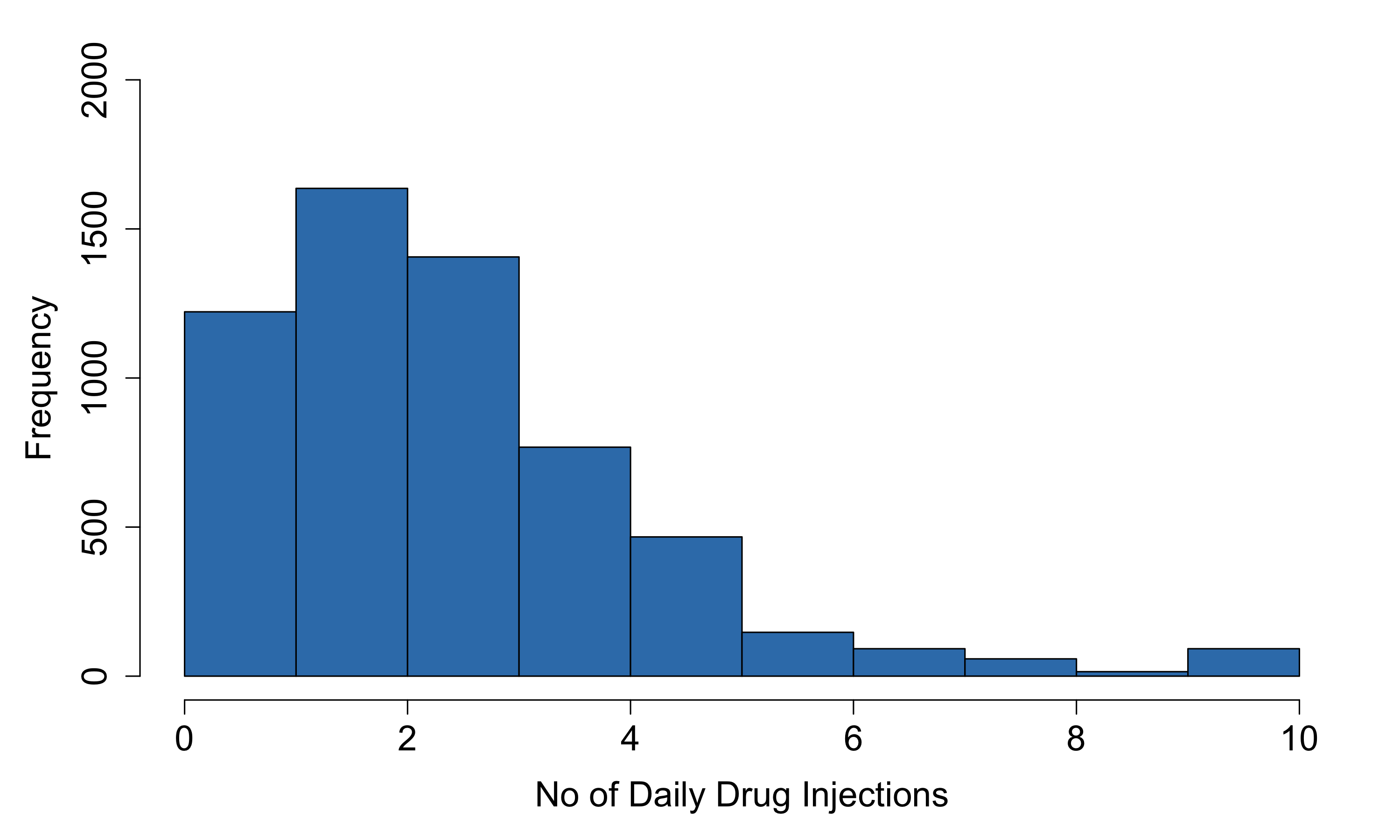}
			\caption{Distribution of the daily drug injections}
			\label{fig:freqDrug}
		\end{figure}
		\item Number of times reusing own syringes in 30 days: \textcolor{black}{\(\mu = 1.61,\  \sigma = 6.15\)}
		\begin{figure}[H]
			\centering
			\includegraphics[width=0.8\textwidth]{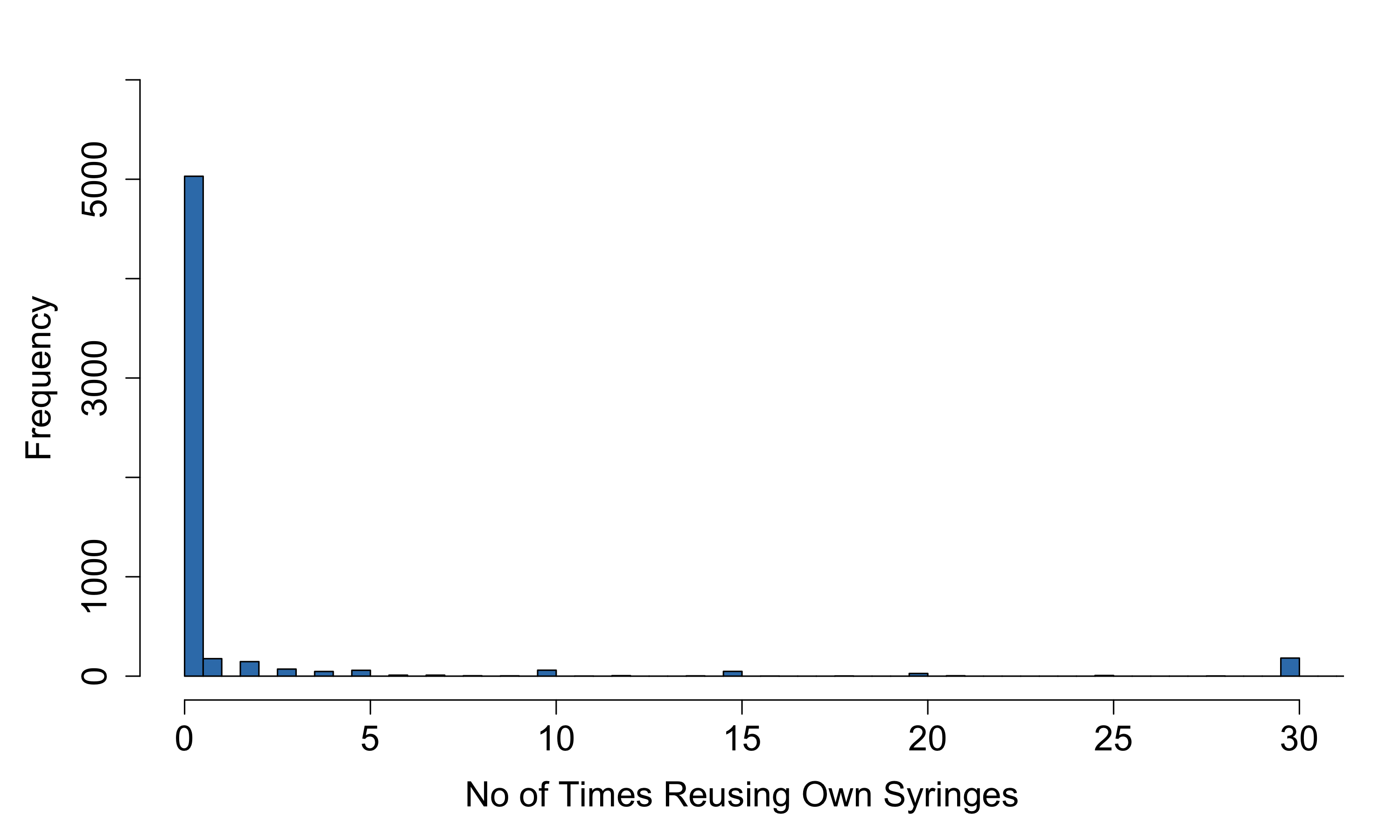}
			\caption{Distribution of the number of times reusing own syringes in 30 days}
			\label{fig:freqReuse}
		\end{figure}
		\item Number of times using others' used syringes in a 30 days: \textcolor{black}{\(\mu = 0.34, \ \sigma = 1.29\)}
		\begin{figure}[H]
			\centering
			\includegraphics[width=0.8\textwidth]{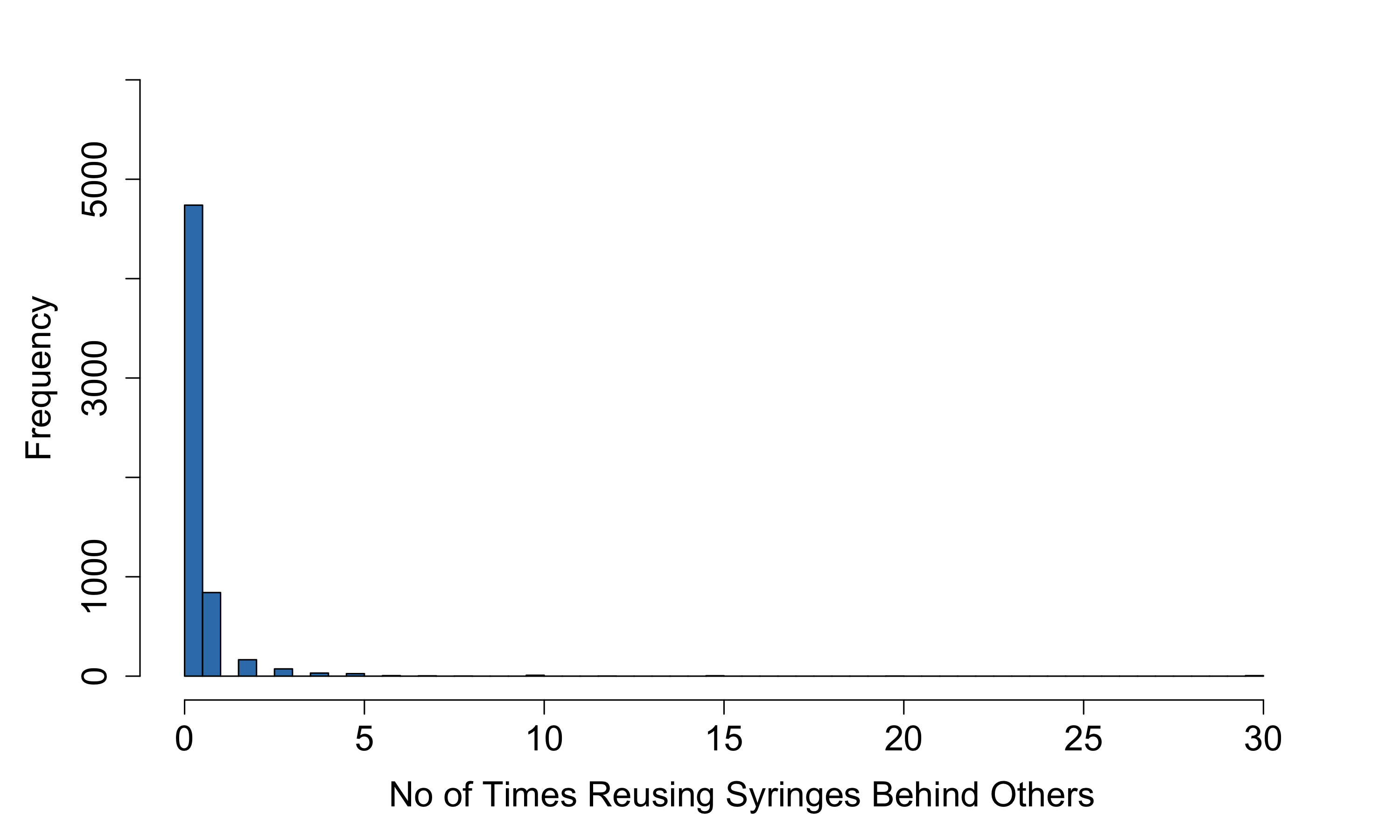}
			\caption{Distribution of the number of times using others' used syringes in a 30 days}
			\label{fig:freqUsebehind}
		\end{figure}
		\item Number of times visiting the area of service locations in 30 days: \(\mu = 23.88, \ \sigma = 9.75\)
		\begin{figure}[H]
			\centering
			\includegraphics[width=0.8\textwidth]{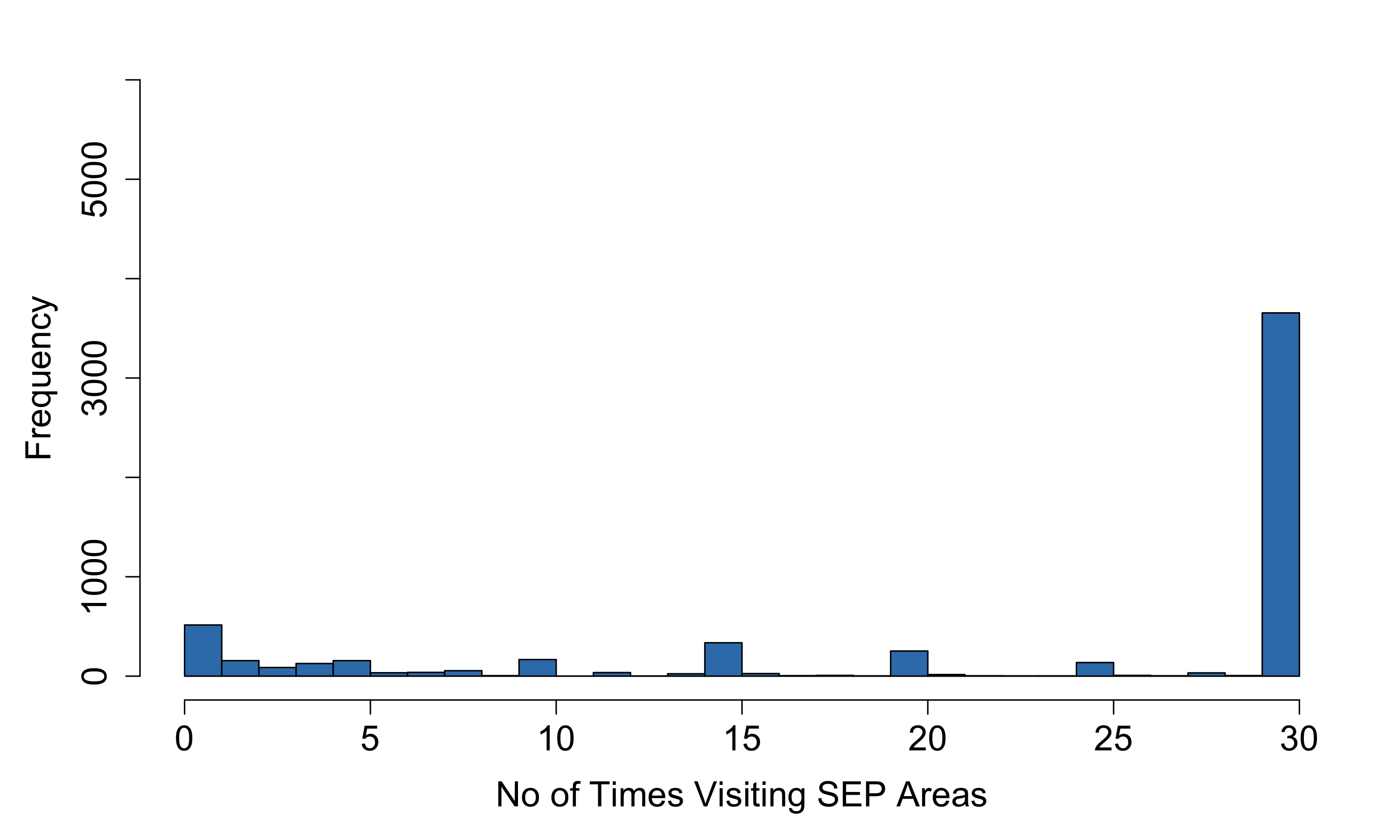}
			\caption{Distribution of the number of times visiting the area of service locations in 30 days}
			\label{fig:freqInarea}
		\end{figure}
	\end{itemize}
	
	\section{Statistical Significance Results of Fitted Parameters} \label{appen:coeffSig}
	We aim to test the significance of each fitted coefficient, by using a bootstrap resampling scheme. We take 100 bootstrap replicates with the same size of the data. For each replicate, we solve the nonlinear optimization model~\eqref{prob:betagammaN} and obtain a set of coefficients. Then we calculate \(\Delta\) and \(T\), as the inter-arrival time change and the expected sojourn time change as described in Section~\ref{subsec:fitting}, for each replicate based on its fitted coefficients. We count how many of those 100 replicates of \(b_{i,j}\), \(g_{i,j}\), \(\rho_j\), \(\Delta_j\) and \(T_j\) are positive. If a large portion of them, like 90\%, is positive or negative, then we can conclude that the parameter is significantly nonzero.
	
	Each column in Table~\ref{table:Sig} shows the number of samples, among 100 samples, of which the fitted coefficient is positive. In addition to acronyms defined in the main text, FUSBO stands for frequency of using syringes behind (i.e., after) others.
	\begin{table}[H]
		\centering
		\begin{tabular}{l | l | l | l | l | l | l | l |}
			\hline
			\(\qquad\)& Factor \(j\) & \(\rho_{j}\) & \(b_{1,j}\) & \(g_{1,j}\) & \(g_{2,j}\) & \(\Delta\) & \(T\) \\ \hline
			&	Snort	&	36	&	16	&	89	&	54	&	67	&	69	\\ \cline{2-8}
			&	Gallery	&	100	&	7	&	57	&	77	&	88	&	1	\\ \hline
			&	From Other Locations	&	97	&	13	&	81	&	96	&	20	&	0	\\ \cline{2-8}
			&	From Other SEP	&	1	&	96	&	7	&	59	&	13	&	98	\\ \cline{2-8}
			&	From Family	&	68	&	55	&	45	&	27	&	73	&	40	\\ \cline{2-8}
			&	From Friends	&	5	&	52	&	26	&	0	&	100	&	100	\\ \cline{2-8}
			&	From Acquaintance	&	41	&	63	&	53	&	45	&	41	&	48	\\ \cline{2-8}
			&	From Strangers	&	9	&	57	&	62	&	93	&	0	&	49	\\ \hline
			&	Speedball	&	19	&	1	&	91	&	4	&	100	&	99	\\ \cline{2-8}
			&	Heroin	&	0	&	15	&	67	&	10	&	96	&	100	\\ \cline{2-8}
			&	Cocaine	&	42	&	54	&	52	&	56	&	34	&	48	\\ \cline{2-8}
			&	Ritalin Heroin	&	29	&	32	&	3	&	13	&	65	&	64	\\ \cline{2-8}
			&	Other Drug	&	62	&	53	&	29	&	37	&	75	&	58	\\ \hline
			&	In Treatment	&	0	&	94	&	0	&	64	&	20	&	99	\\ \cline{2-8}
			&	Been in Treatment	&	9	&	10	&	90	&	32	&	88	&	98	\\ \cline{2-8}
			&	Attempted Treatment	&	41	&	30	&	83	&	7	&	95	&	84	\\ \cline{2-8}
			&	Want Treatment	&	73	&	21	&	4	&	22	&	100	&	74	\\ \hline
			&	Female	&	45	&	77	&	17	&	93	&	8	&	7	\\ \cline{2-8}
			&	Male	&	100	&	75	&	25	&	85	&	20	&	20	\\ \cline{2-8}
			&	Transsexual	&	99	&	99	&	99	&	99	&	99	&	15	\\ \hline
			&	White	&	100	&	96	&	46	&	80	&	2	&	2	\\ \cline{2-8}
			&	African American	&	100	&	19	&	30	&	64	&	70	&	2	\\ \cline{2-8}
			&	Puerto Rican	&	0	&	60	&	99	&	68	&	12	&	100	\\ \cline{2-8}
			&	Mexican	&	0	&	42	&	64	&	99	&	1	&	82	\\ \cline{2-8}
			&	Other Latino	&	27	&	63	&	0	&	44	&	77	&	82	\\ \cline{2-8}
			&	Other	&	10	&	82	&	12	&	43	&	50	&	91	\\ \hline
			&	Age	&	44	&	57	&	7	&	89	&	16	&	55	\\ \cline{2-8}
			&	Age of First Drug Use	&	52	&	100	&	68	&	100	&	0	&	41	\\ \cline{2-8}
			&	Drug Use Span	&	60	&	6	&	2	&	23	&	100	&	43	\\ \cline{2-8}
			&	FUD	&	65	&	98	&	15	&	5	&	55	&	31	\\ \cline{2-8}
			&	FROS	&	91	&	12	&	54	&	61	&	78	&	19	\\ \cline{2-8}
			&	FUSBO	&	34	&	71	&	35	&	63	&	26	&	50	\\ \cline{2-8}
			&	FBSA	&	0	&	51	&	100	&	99	&	0	&	100	\\ \hline
		\end{tabular}
		\caption{Statistics of bootstrap samples for estimating coefficients \(\rho\), \(b\), and \(g\), as well as changes to the mean time to return to an SEP site, \(\Delta\), and the expected sojourn time in the system, \(T\)}
		\label{table:Sig}
	\end{table}
\end{document}